\documentclass[aps,pre,groupedaddress,showpacs,preprint]{revtex4}
\usepackage{graphicx}
\usepackage{dcolumn}
\usepackage{subfigure}%
\usepackage{bm}
\usepackage{amssymb}
\usepackage{amsmath}
\usepackage{cases}
\usepackage{color,soul}
\usepackage{verbatim} 

\usepackage[latin1]{inputenc}   

\usepackage{ulem}

\usepackage{color,soul}  
\setstcolor{red}         

\newcommand{\be}{\begin{equation}}
\newcommand{\ee}{\end{equation}}
\newcommand{\ba}{\begin{eqnarray}}
\newcommand{\ea}{\end{eqnarray}}
\newcommand{\besu}{\begin{subequations}}
\newcommand{\esu}{\end{subequations}}

\begin{document}
\title{Continuous-time random-walk model for anomalous diffusion in expanding media}
\author{F. Le Vot$^{1}$, E. Abad$^{2}$, S. B. Yuste$^{1}$}
\affiliation{
 $^{1}$ Departamento de F\'{\i}sica and Instituto de Computaci\'on Cient\'{\i}fica Avanzada (ICCAEX) \\
 Universidad de Extremadura, E-06071 Badajoz, Spain \\
 $^{2}$ Departamento de F\'{\i}sica Aplicada and Instituto de Computaci\'on Cient\'{\i}fica Avanzada (ICCAEX) \\ Centro Universitario de M\'erida \\ Universidad de Extremadura, E-06800 M\'erida, Spain
}

\begin{abstract}
Expanding media are typical in many different fields, e.g. in Biology and Cosmology. In general, a medium expansion (contraction) brings about dramatic changes in the behavior of diffusive transport properties such as the set of positional moments and the Green's function. Here, we focus on the characterization of such effects when the diffusion process is described by the Continuous Time Random Walk (CTRW) model. As is well known, when the medium is static this model yields anomalous diffusion for a proper choice of the probability density function (pdf) for the jump length and the waiting time, but the behavior may change drastically if a medium expansion is superimposed on the intrinsic random motion of the diffusing particle. For the case where the jump length and the waiting time pdfs are long-tailed, we derive a general bifractional diffusion equation which reduces to a normal diffusion equation in the appropriate limit. We then study some particular cases of interest, including L\'evy flights and subdiffusive CTRWs. In the former case, we find an analytical exact solution for the Green's function (propagator). When the expansion is sufficiently fast, the contribution of the diffusive transport becomes irrelevant at long times and the propagator tends to a stationary profile in the comoving reference frame. In contrast, for a contracting medium a competition between the spreading effect of diffusion and the concentrating effect of contraction arises. In the specific case of a subdiffusive CTRW in an exponentially contracting medium, the latter effect prevails for sufficiently long times, and all the particles are eventually localized at a single point in physical space. This ``Big Crunch'' effect, totally absent in the case of normal diffusion, stems from inefficient particle spreading due to subdiffusion.
We also derive a hierarchy of differential equations for the moments of the transport process described by the subdiffusive CTRW model in an expanding medium. From this hierarchy, the full time evolution of the second-order moment is obtained for some specific types of expansion. In the case of an exponential expansion, exact recurrence relations for the Laplace-transformed moments are obtained, whence the long-time behavior of moments of arbitrary order is subsequently inferred.
Our analytical and numerical results for both Lévy flights and subdiffusive CTRWs confirm the intuitive expectation that the medium expansion hinders the mixing of diffusive particles occupying separate regions. In the case of L\'evy flights, we quantify this effect by means of the so-called ``L\'evy horizon''.

\end{abstract}


\maketitle

\section{Introduction}

There are numerous examples in Biology and Cosmology where stochastic transport takes place in an expanding medium. Among other processes, Fickian diffusion models have been used to study cell migration in growing tissues \cite{Binder, Landman, Simpson2015c} and cosmic ray diffusion in the extragalactic space \cite{Berezinsky2006, Berezinsky2007, Kotera}. Anomalous diffusion, on the other hand, is a common phenomenon in biological systems \cite{Metzler2014b}, and has also been argued to be relevant for the propagation of cosmic rays \cite{Lagutin2003, Uchaikin2013}. In such cases, mesoscopic random walk models provide a very useful theoretical rationale to understand the observed behavior. This highlights the importance of developing a suitable random walk description accounting for the effect of the medium expansion.

In order to contribute to the above task, our starting point is a recent article \cite{Yuste2016} where the differential equation for Brownian diffusion in an expanding medium was derived via a mesoscopic Chapman-Kolmogorov approach. The behavior of some characteristic properties such as the hierarchy of positional moments or the associated Green's function were studied by expressing the physical distance in terms of the comoving distance. In this follow-up work we continue to exploit this idea to investigate the behavior of a broad class of Continuous Time Random Walk (CTRW) models \cite{Kutner2017} by means of the associated fractional diffusion equation (FDE) \cite{Metzler2000, Soko2012}. This class of CTRW models includes the cases where either the walker's jump length pdf, the walker's waiting time pdf or both of these quantities display an asymptotic power-law behavior. Such long tails are known to generate anomalous diffusion when the medium is static \cite{Klafter2011}, and the question is how the observed behavior is modified if one allows for a medium expansion. We anticipate that the phenomenology of the transport process changes dramatically if the expansion is sufficiently fast.  A strong modification of the transport properties also takes place in the opposite scenario of a fast contracting medium, a situation in which our FDE remains applicable. Here, we shall restrict ourselves to the case of a one dimensional system subject to a uniform expansion (the derivation of the relevant FDE in higher dimensions is straightforward).

The remainder of this paper is organized as follows. In Sec.~\ref{secII}, we derive a bifractional FDE which constitutes the cornerstone of our subsequent analysis. In Sec.~\ref{secIII}, we address the specific case of L\'evy flights \cite{Metzler2007}, a superdiffusive process widely used to model very diverse processes, e.g., foraging behavior \cite{Atkinson2002,Ramos2004}, spreading of diseases \cite{Hufnagel2004}, hopping processes along polymers \cite{Sokolov1997}, or even the trajectories of one-dollar bills  \cite{Brockmann2006}. In Sec.~\ref{secIV}, we study the subdiffusive CTRW in an expanding medium. One of the first applications of the subdiffusive CTRW model was the study of charge carrier transport in amorphous semiconductors \cite{Scher1973a, Scher1973b}, but it has subsequently been applied to many other systems (see e.g. the recent references \cite{Kutner2017,Barzykin1994, Berkowitz2006, LeBorgne2011, Hornung2005, YusteAbadLindenberg2010, FedotovFalconer2014}).
In a biological context, taking into account the medium expansion might help to refine the existing CTRW models for certain phenomena such as morphogen gradient formation, where tissue growth is known to play a role during the dispersion of the morphogens \cite{Fried2014,Averbukh2014}.
Finally, in Sec.~\ref{Conclusions} we summarize our main results and briefly outline the possible implications of the latter.

\section{CTRW model in an expanding medium: derivation of the FDE}
\label{secII}

The traditional CTRW model \cite{MontrollWeiss} consists of a particle (the ``walker'') performing instantaneous jumps of random length $\Delta y$ after randomly distributed waiting times $\Delta t$. The walker's motion is described by a joint pdf $\psi(\Delta y,\Delta t)$ associated with the probability $\psi(\Delta y,\Delta t) d\Delta y d\Delta t$ for the walker to perform a jump of length between $\Delta y$ and $\Delta y+d\Delta y$ after a waiting time between $\Delta t$ and $\Delta t+d\Delta t$. In the general situation, both random variables $\Delta y$ and $\Delta t$ are coupled \cite{Klafter2011}, but for convenience we shall restrict ourselves here to the standard assumption of an uncoupled CTRW, implying that the following factorization holds:
 \begin{equation}
\psi (\Delta y,\Delta t) = \varphi(\Delta t) \lambda(\Delta y).
\label{DecouplementCTRW}
\end{equation}
Clearly, $\varphi(\Delta t) d\Delta t $ gives the probability that the waiting time between two consecutive walker steps lies in the interval $[\Delta t,\Delta t+d\Delta t)$ regardless of the step size, whereas $\lambda(\Delta y) d\Delta y$ is the probability that the jump length takes a value between $\Delta y$ and $\Delta y+d\Delta y$. For simplicity, we shall restrict ourselves here to the case of symmetric distributions $\lambda(\Delta y)$.

Let us now assume that a medium expansion takes place. In addition to the physical coordinate $y$ (related to the proper displacement of the particle), it is convenient to introduce a comoving coordinate $x$ associated with a reference frame where the expanding medium appears to be static. Loosely speaking, in what follows we shall occasionally term ``physical quantities'' those quantities referring to the physical coordinate and ``comoving quantities'' those quantities referring to the comoving coordinate.

An informal, yet accurate way of defining the comoving coordinate of a particle which goes beyond the 1D case is as follows. To this end, the medium can be thought of as a continuum of ``fixed points'' lacking intrinsic motion, although these fixed points are of course drifted away by the expansion of the medium (or, in the language of Cosmology, by the so-called Hubble flux).  It is clear that at any time $t$ the walker occupies the position of a fixed point. The position of this fixed point at the initial time $t_0$ is the comoving coordinate of the walker at any time $t$. Obviously, the comoving and the physical frames of reference are equal at the initial time.

Focusing now on a 1D system, in the case of a uniform expansion, both coordinates are related to one another in a very simple way, i.e., \cite{Yuste2016}
\begin{equation}
y = a(t) x \quad \mbox{ with } \quad a(t_0)=1.
\label{PhysComRel}
\end{equation}
In passing, we note that in the cosmological context of an expanding universe, the function $a(t)$ is the so-called scale factor \cite{Peacock1999, Ryden2003}.

In practice, a computational realization of the above random walk process in physical space could be carried out by concatenating events consisting of a medium expansion and a subsequent particle jump. Each of such events would be implemented as follows. First, a value for the waiting time $\Delta t$ would be drawn by using the corresponding pdf $\varphi(\Delta t)$. Then the physical space metric would be stretched by a factor $a(t)$. Finally, an instantaneous jump would be implemented by using the pdf $\lambda(\Delta y)$ to generate a value for the jump length $\Delta y$. Thus, the medium expansion between consecutive jumps has no effect on the comoving coordinate.  This is in fact the procedure we employ to carry out our simulations (see details in the Appendix).

Note that the jump length pdf $\lambda(\Delta y)$ describes the intrinsic random motion of the particle, and it does \emph{not} depend on time. Then, it immediately follows from Eq.~\eqref{PhysComRel} that the distribution $\lambda (\Delta x)$ describing the jump length in comoving coordinates should be time dependent. In order to make this time dependence explicit in our notation, we shall write this pdf as $\lambda(\Delta x|t)$.  Clearly, one has
\begin{equation}
\lambda(\Delta x|t) = \lambda(\Delta y) \frac{dy}{dx} = a(t)\lambda(\Delta y),
\label{LambdaProperToComoving}
\end{equation}
where $t$ denotes the instant when the walker jumps and $\Delta y=a(t) \Delta x$. In order to obtain a description of the random walk in the comoving reference frame, Eq.~(\ref{DecouplementCTRW}) must be replaced with the joint pdf
\begin{equation}
\psi(\Delta x|t,t-t') = \varphi(t-t') \lambda(\Delta x|t),
\end{equation}
which gives the probability to perform a jump of comoving length $\Delta x$ at time $t$ if the previous jump took place at time $t'<t$.

A standard choice for $\lambda(\Delta y)$ is a Gaussian pdf. To simplify the resulting expressions, we denote the second moment by $2\sigma^2$ \cite{Metzler2000}, i.e.,
\begin{equation}
\label{gaussianLambda1}
\lambda(\Delta y) = \frac{1}{\sqrt{4 \pi \sigma^2}} \exp \left( \frac{-\Delta y^2}{4 \sigma^2} \right),
\end{equation}
leading to
\begin{equation}
\label{gaussianLambda2}
\lambda(\Delta x | t) = \frac{1}{\sqrt{4 \pi \sigma^2 / a^2(t)}} \exp \left( \frac{-\Delta x^2}{4 \sigma^2 / a^2(t)} \right)
\end{equation}
in terms of the comoving coordinate.
With this choice of Gaussian distributed displacements --or, analogously, for any other distribution whose second moment is finite--, the walker's type of diffusion will be determined by the temporal part of $\psi(\Delta y, \Delta t)$. Normal diffusion is e.g. obtained in the case of an exponential waiting time pdf
\begin{equation}
\label{expphi}
\varphi(\Delta t) = \frac{1}{\tau} \exp \left( - \frac{\Delta t}{\tau} \right),
\end{equation}
or any other pdf with finite average waiting time $\tau$.  In contrast, for a long-tailed pdf displaying the asymptotic long-time behavior
\begin{equation}
\label{longtailphi}
\varphi (\Delta t) \sim \frac{\alpha}{\Gamma(1-\alpha)} \frac{\tau^{\alpha}}{\Delta t^{1 + \alpha}},
\end{equation}
one obtains anomalous diffusion in the regime $0 < \alpha < 1$. In this case, a mean waiting time does not exist, and consequently $\tau$ is only a typical time scale for the asymptotic long-time decay of $\varphi (\Delta t)$ for a fixed value of $\alpha$. This class of pdfs is exemplified by Pareto or Pareto-like distributions \cite{Foss2011} as well as by pdfs given by the derivative of the Mittag-Leffler function \cite{Heinsalu2006} or by the family of one-sided L\'evy stable distributions \cite{Condamin2007,Klafter2011}.

As far as $\lambda(\Delta y)$ is concerned, an important special case is given by the power-law pdf
\begin{equation}
\lambda(\Delta y) \sim \sin \left( \frac{\pi \mu}{2} \right) \frac{ \Gamma (1+\mu)}{\pi} \sigma^\mu |\Delta y|^{-1-\mu}, \quad \sigma, \mu >0
\label{longtailLambda}
\end{equation}
for large values of $|\Delta y|$. The choice $0<\mu<2$ corresponds to a so-called L\'evy flight, which provides a minimal model for a broad class of random motions in Nature, notably in the context of ecology (target search processes, foraging, predator-prey models, etc.). In many of such instances, the target search process consists of long jumps aimed at exploring distant environments followed by a succession of short jumps for a detailed local search.

Let us now focus on the case where $\varphi(\Delta t)$ and $\lambda(\Delta y)$ are both long-tailed, that is, they are respectively given by Eqs.~\eqref{longtailphi} and \eqref{longtailLambda}. Our goal here will be to obtain the diffusion equation for the sojourn pdf $P(y,t)$ of the particle in physical space, that is, $P(y,t) dy$ is the probability to find a walker starting from the origin inside the infinitesimal interval $[y,y+dy)$ at time $t$. To this end, it is convenient to switch to comoving coordinates, since in this reference frame the position of the walker does not change as long as she does not jump. This will allow us to employ the standard formalism for the CTRW model in a static medium.

Let us hereafter denote by $W(x,t)$ the comoving counterpart of $P(y,t)$, that is, $W(x,t)$ is the probability density to find the particle at position $x$ at time $t$. Both pdfs are related to one another as follows:
\begin{equation}
W\left(x,t\right)=a(t)\,P(y,t).
\end{equation}
Following now a well-trodden path, let us introduce $\eta(x,t)$ as the pdf associated with the event of arriving at position $x$ exactly at time $t$. As is the case in the absence of the expansion, this arrival pdf  fulfils the master equation \cite{Klafter2011,Metzler2000}
\begin{equation}
\label{arrivaldensEq}
\eta(x,t) = \int_{-\infty}^{\infty} dx' \int_0^t dt' \eta(x',t') \psi(x-x'|t, t-t') + \delta(t) \delta(x).
\end{equation}
The second term of the right-hand side accounts for the initial condition. The first term comes from the probabilistic description of the fact that if a particle arrived at position $x$ at time $t$, it may have arrived at any position $x'$ at a previous time $t'$, and then have spent a time $t-t'$ before moving from $x'$ to $x$.

In turn,  $W(x,t)$ is given by the equation
\begin{equation}
W(x,t)=\int_0^t dt' \eta(x,t') \int_{t-t'}^\infty \varphi(t'') dt''.
\label{comovPropIntEq}
\end{equation}
This equation simply states that the probability to find the walker in the vicinity of a given position $x$ at time $t$ is obtained by accumulation of all the possible events in which the arrival of the walker at $x$ took place by an instantaneous jump at some previous time $t'$, with no further jumps occurring between $t'$ and $t$.

We are now in the position to apply standard arguments and thereby obtain a FDE from Eqs.~\eqref{arrivaldensEq} and \eqref{comovPropIntEq}. Switching to Fourier-Laplace space, Eq.~\eqref{comovPropIntEq} becomes
\begin{equation}
\widehat{\tilde{W}}(k,s)=\widehat{\tilde{\eta}}(k,s) \frac{1-\tilde{\varphi}(s)}{s}.
\label{FourierLapProp}
\end{equation}
The counterpart of Eq.~\eqref{arrivaldensEq} in Fourier space then reads as follows:
\begin{equation}
\hat{\eta}(k,t) =  \hat{\lambda}(k|t) \int_0^t dt' \hat{\eta}(k,t') \varphi(t-t') + \delta(t),
\label{EtaInFourierSpace}
\end{equation}
where, from Eq.~\eqref{LambdaProperToComoving}, the Fourier transform of the comoving jump length pdf is simply
\begin{equation}
 \hat{\lambda}(k|t)=\hat{\lambda}(k/a(t)).
 \label{FTrelation}
 \end{equation}
Here, $ \hat{\lambda}(k|t) \equiv \mathcal{F}[\lambda(\Delta x|t)]$  and  $\hat{\lambda}(k) \equiv \mathcal{F}[\lambda(\Delta y)]$.
Taking the Laplace transform of Eq.~\eqref{EtaInFourierSpace} we find
\begin{equation}
\widehat{\tilde{\eta}}(k,s)=\mathcal{L}\left\{
\hat{\lambda}(k|t) \mathcal{L}^{-1}
\left[
\widehat{\tilde{\eta}}(k,s) \tilde{\varphi}(s)
\right]
\right\} +1.
\label{FourierLapEta}
\end{equation}

In order to advance further, detailed knowledge of the jump length pdf and the waiting time pdf is needed. Focusing now on the special case of L\'evy flights, the Fourier-transform of the jump length pdf \eqref{longtailLambda} displays the well-known behavior
\begin{equation}
\label{lambdak}
 \hat{\lambda}(k) \sim 1 - \sigma^\mu |k|^\mu
\end{equation}
for $k\to 0$. According to Eq.~\eqref{FTrelation}, one then has
\begin{equation}
 \hat{\lambda}(k|t) \sim 1 -  \sigma^\mu |k|^\mu/a^\mu(t).
\end{equation}
Inserting this relation into Eq.~\eqref{FourierLapEta} and taking Eq.~\eqref{FourierLapProp} into account, one obtains
\begin{equation}
s \widehat{\tilde{W}}(k,s)-1= \sigma^{\mu} |k|^{\mu}\mathcal{L} \left\{ \frac{1}{a^{\mu}(t)} \mathcal{L}^{-1} \left[ \frac{s \tilde{\varphi}(s)}{1 -  \tilde{\varphi}(s)}  \widehat{\tilde{W}}(k,s)  \right]  \right\} .
\label{FourierLapProp2}
\end{equation}

 We now consider the case where the long time behavior of $\varphi$ is given by the long-tailed pdf \eqref{longtailphi}. In the $s\to 0$ limit, the Laplace transform behaves as follows:
 \begin{equation}
 \tilde{\varphi}(s)\sim 1-\tau^\alpha s^\alpha, \qquad s\to 0,
 \label{varphiAsin}
\end{equation}
 and, consequently,
 \begin{equation}
 s \tilde{\varphi} / ( 1-\tilde{\varphi}) \sim s^{1-\alpha} / \tau^{\alpha}.
 \end{equation}
 Inserting this asymptotic form into Eq.~\eqref{FourierLapProp2}, one is left with the equation
 \begin{equation}
 s \widehat{\tilde{W}}(k,s)-1=-|k|^\mu \frac{\sigma^\mu}{\tau^\alpha} \mathcal{L}\left\{
\frac{1}{a^\mu(t)} \mathcal{L}^{-1}\left[s^{1-\alpha}\widehat{\tilde{W}}(k,s)\right]
\right\}.
\label{FTPropEq}
\end{equation}

The next step is to take the inverse Fourier-Laplace transform of the above equation. The structure of Eq.~\eqref{FTPropEq} is inherited from the asymptotic power law behavior of the Fourier transformed jump length pdf and of the Laplace transformed waiting time pdf, respectively described by Eqs.~\eqref{lambdak} and \eqref{varphiAsin}. Ultimately, this behavior gives rise to expressions where the Fourier (Laplace) transform of the relevant function is  multiplied by  the absolute value of  the wave vector $k$  (Laplace variable, $s$) raised to a positive exponent (see below). On the other hand, the inverse transforms of such expressions are known to be given by non-local integrodifferential operators (so-called fractional derivatives). The type of fractional derivative is determined both by the allowable range of the relevant variable (in our case, position or time) and by the range of the exponent describing its long-tail behavior \cite{Metzler2000}.
 Assuming that $f(t)$ is a generic function of the time variable ($t\ge 0$) and using the short-hand notation $\tilde{f}(s)$ for its Laplace transform $\mathcal{L}[f(t)]$, one has \cite{Podlubny1999}
\begin{equation}
\mathcal{L} \{
  {_{0}}\mathcal{D}_t^{1-\alpha} f(t)\}=s^{1-\alpha} \tilde{f}(s),
\end{equation}
where ${_{0}}\mathcal{D}_t^{1-\alpha} f(t)$ stands for the Gr\"{u}nwald-Letnikov fractional derivative of $f(t)$. If $f(t)$ is a sufficiently smooth function at the origin so that the condition
$
\lim_{t\to 0} \int_0^t (t-\tau)^{\alpha -1} f(\tau)d\tau=0
$
is fulfilled, one has
$
  {_{0}}\mathcal{D}_t^{1-\alpha} f(t)=\,_{~~0}^{{RL}}D_t^{1-\alpha} f(t),
$
(see e.g. Eqs.~(2.255), (2.248) and (2.240) in \cite{Podlubny1999})
where
\begin{equation}
\,_{~~0}^{{RL}}D_t^{1-\alpha} f(t)=\frac{1}{\Gamma(\alpha)} \frac{d}{dt}
\int_0^t \, dt'\; \frac{f(t')}{(t-t')^{1-\alpha}}
\end{equation}
defines the so-called Riemann-Liouville fractional derivative.

Next, let us consider a generic function $f(x)$ in the spatial variable $x$ ($-\infty <x<\infty$), and
define an operator $\nabla^{\mu}$ via the equation
\begin{equation}
{\mathcal F}[\nabla^{\mu}f(x)]=-|k|^\mu \hat{f}(k),
\label{RFdef}
\end{equation}
where the short-hand notation $\hat f(k)$ for the Fourier transform ${\mathcal F}[f(x)]$ has been used. It turns out that this operator is the so-called  Riesz-Weyl fractional derivative $\nabla^{\mu} f(x)$ \cite{Klages2008}, which can be explicitly expressed as follows:
\begin{equation}
\nabla^{\mu} f(x) = \Gamma (1+\mu)\, \frac{\sin (\mu \pi/2)}{\pi} \int_0^{\infty} d\xi \, \frac{f(x+\xi) -2 f(x) + f(x-\xi) }{\xi^{1+\mu}}
\end{equation}
for $0<\mu<2$ \cite{Mainardi01}.

We are now in the position to express the inverse Fourier Transform of the rhs of Eq.~\eqref{FTPropEq} in terms of fractional derivatives and thereby arrive at our main result, i.e., the bifractional differential equation
\begin{equation}
\label{bifractionaleq}
\frac{\partial W (x,t)}{\partial t} = \frac{\mathfrak{D}_{\alpha, \mu}}{a^{\mu}(t)}\, \nabla^{\mu}\,  {_0}\mathcal{D}_t^{1-\alpha} W(x,t) .
\end{equation}
Here, the quantity
\begin{equation}
\mathfrak{D}_{\alpha, \mu} = \sigma^{\mu} / \tau^{\alpha}
\label{diffconst}
\end{equation}
is the so-called anomalous diffusion constant.

For a waiting time pdf with a finite mean, $_0\mathcal{D}_t^{1-\alpha}$ becomes the identity operator ($\alpha\to 1$) and Eq.~\eqref{bifractionaleq} reduces to the following FDE:
\begin{equation}
\frac{\partial W (x,t)}{\partial t} = \frac{\mathfrak{D}_{1, \mu}}{a^{\mu}(t)} \,\nabla^{\mu} W(x,t) .
\label{ComovingNormalLevyFligthFDE}
\end{equation}
Similarly, for a jump length pdf with a finite second moment, the operator $\nabla^\mu$ becomes the usual Laplacian operator ($\mu\to 2$), and one obtains the following equation:
\begin{equation}
\frac{\partial W (x,t)}{\partial t} = \mathfrak{D}_{\alpha} \frac{1}{a^2(t)} ~ \frac{\partial^2 }{\partial x^2}\, \,  {_0}\mathcal{D}_t^{1-\alpha}\,  W(x,t),
\label{AnomalousExpansiveDiffusionEquation}
\end{equation}
where $\mathfrak{D}_{\alpha} \equiv \mathfrak{D}_{\alpha,2}=\sigma^2 / \tau^{\alpha}$. Of course, if both limiting cases occur simultaneously, Eq.~\eqref{bifractionaleq} becomes the standard diffusion equation for homogeneously expanding media \cite{Yuste2016}.

\section{L\'evy flights in expanding media}
\label{secIII}

L\'evy flights can be modeled by a CTRW with a long-tailed jump length pdf (\ref{longtailLambda}) and a waiting time pdf with finite mean. When L\'evy processes take place in expanding media, the pdf $W(x,t)$ satisfies Eq.~\eqref{ComovingNormalLevyFligthFDE}, and the associated diffusion properties will be manifestly different from the ones for the case of a static medium. The present section is devoted to a detailed discussion of the new phenomenology. One interesting feature is that the probability that two diffusing particles with a sufficiently large initial separation meet may become negligibly small. This behavior is confirmed both analytically and numerically.

\subsection{Propagator}

The solution of Eq.~\eqref{ComovingNormalLevyFligthFDE} is greatly simplified by introducing  the diffusion conformal time,
\begin{equation}
T(t) = \int_0^t \frac{ds}{a^{\mu} (s)}.
\label{ConformalTime}
\end{equation}
With this change of variable, Eq.~\eqref{ComovingNormalLevyFligthFDE} is reduced to the standard fractional differential equation for a L\'evy flight, that is,
\begin{equation}
\frac{\partial W (x,T)}{\partial T} = \mathfrak{D}_{1, \mu} \nabla^{\mu} W(x,T).
\label{LFdiffeq}
\end{equation}
Note that Eq.~\eqref{LFdiffeq} has the same form as the equation corresponding to the standard case of a static medium, except for the fact that the time $t$ is replaced by the conformal time $T$. While the behavior of $T$ as a function of $t$ depends on the specific type of medium expansion and may in general give rise to new physics, the standard tools for the solution of diffusive problems can still be applied. In the following, we shall use these techniques to obtain the propagator   associated with Eq.~\eqref{LFdiffeq} [recall that the propagator or Green's function is defined as the solution in full space ($-\infty<x<\infty$) that corresponds to a Dirac delta initial condition, i.e., $W(x,T=0)=\delta(x)$]. To this end, we first note that the Fourier transform of Eq.~\eqref{LFdiffeq} is
\begin{equation}
\frac{\partial\widehat W (k,T)}{\partial T}=- \mathfrak{D}_{1, \mu} |k|^\mu \widehat W(k,T),
\end{equation}
where we have made use of Eq.~\eqref{RFdef}.  The solution of the above equation satisfying the Dirac delta initial condition is
\begin{equation}
\widehat W (k,T)=\exp\left(-\mathfrak{D}_{1, \mu} |k|^\mu T\right),
\label{solFE}
\end{equation}
which turns out to be the characteristic function of the symmetric L\'evy function with decay exponent (or L\'evy index) $\mu$ and scale factor $\sigma=(\mathfrak{D}_{1,\mu} T)^{{1}/{\mu}}$,
hereafter denoted by
$\mathsf{L}_{\mu} \left( x;\mathfrak{D}_{1,\mu} T\right)$, i.e.,
\begin{equation}
 W (x,T)=\mathsf{L}_{\mu} \left( x;\mathfrak{D}_{1,\mu} T \right).
\label{solFExT}
\end{equation}
Note that for typographical simplicity, we use the notation $\mathsf{L}_{\mu}(x;\sigma^\mu)$ instead of the more usual notation $\mathsf{L}_{\mu,\sigma}(x)=\mathsf{L}_{\mu,\beta,\gamma,\sigma}(x)$ with $\beta=\gamma=0$,  $\beta$ being the asymmetry parameter and $\gamma$ the shift parameter \cite{Klafter2011}.
Note that
taking into account the scaling property \cite{Bouchaud1990}
\begin{equation}
\label{LevyScaling}
 \mathsf{L}_{\mu}(x;C)=C^{-1/\mu} \, \mathsf{L}_{\mu}(x C^{-1/\mu};1),
\end{equation}
the L\'evy function  $\mathsf{L}_{\mu}(x;\sigma^\mu)$ can be rewritten in terms of the normalized L\'evy function $\mathsf{L}_{\mu}(x;1)=\mathsf{L}_{\mu}(x)$ as  $\sigma^{-1}\mathsf{L}_{\mu}(x/\sigma)$.
An alternative representation of the Green's function   is  \cite{Mainardi2005}:
\begin{subequations}
\begin{itemize}
\item  For   $0 < \mu < 1$:
\begin{align}
W (x,T)&=  \frac{1}{\mu (\mathfrak{D}_{1, \mu} T)^{1/\mu}} H^{1,1}_{2,2} \left[ \frac{(\mathfrak{D}_{1, \mu} T)^{\frac{1}{\mu}}}{|x|} \left| \begin{array}{lc}
            (1,1), &    \left(\frac{1}{2}, \frac{1}{2} \right) \\
            (\frac{1}{\mu},\frac{1}{\mu}), &  \left(\frac{1}{2}, \frac{1}{2} \right) \\
             \end{array}
   \right. \right]
\end{align}
\item For $\mu = 1$:
\begin{align}
W (x,T)&= \frac{1}{\pi} \frac{\mathfrak{D}_{1, 1} T}{x^2 + (\mathfrak{D}_{1, 1} T)^2}
\end{align}
\item For $1< \mu < 2$:
\begin{align}
W (x,T)&= \frac{1}{\mu |x|} H^{1,1}_{2,2} \left[ \frac{|x|}{(\mathfrak{D}_{1, \mu} T)^{\frac{1}{\mu}}} \left| \begin{array}{lc}
            (1, \frac{1}{\mu}), &    (1, \frac{1}{2} ) \\
            (1,1), &  (1, \frac{1}{2} ) \\
             \end{array}
   \right. \right],
\end{align}
\end{itemize}
\label{ComovingNormalLevyFligthPropagator}
\end{subequations}
where $H^{1,1}_{2,2}(\cdot)$ denotes Fox's H function \cite{Mathai1978, Mainardi2005, Metzler2000}.

\subsection{Fractional moments}

A standard way of characterizing the particle spreading due to a diffusion process is to evaluate the moments of its propagator, and the mean square displacement $\langle x^2\rangle$ in particular. However, for L\'evy flights in an infinite medium neither the second moment nor higher order moments exist and one has to resort to other approaches (see Sec.~\ref{secwidthPropagator}). For example, a typical quantity characterizing the diffusive spread of L\'evy flights is the $\nu$-th order fractional moment of the propagator $\langle |x|^{\nu} \rangle$ where $0<\nu<\mu \leq 2$ \cite{Metzler2000}.
From Eq.~\eqref{ComovingNormalLevyFligthPropagator} one can find the fractional $\nu$th moment for L\'evy flights in a homogeneous expanding medium for any scale factor $a(t)$:
\begin{equation}
\langle |x|^{\nu} \rangle =\int_{-\infty}^\infty dx\, |x|^{\nu} W(x,t)=   \left( C_{\nu\mu}\, \mathfrak{D}_{1, \mu} \, T\right)^{\frac{\nu}{\mu}}
\label{ComovingNormalLevyFligthFractionalMoment}
\end{equation}
with
\begin{equation}
C_{\nu\mu}= \left[ \frac{2}{\mu} \,\frac{\Gamma \left(-\frac{\nu}{\mu} \right) \Gamma (1 + \nu ) }{\Gamma \left(-\frac{\nu}{2} \right) \Gamma \left(1 + \frac{\nu}{2} \right)} \right]^{\mu/\nu}.
\label{Cnumu}
\end{equation}
This result is obtained from Eq.~\eqref{ComovingNormalLevyFligthPropagator} by using the following property of the H function \cite{Mathai1978, Metzler2000}:
\begin{equation}
\int_0^{\infty} dz ~ z^{\nu-1}  H^{m,n}_{p,q} \left[ c z \left| \begin{array}{lc}
            (a_p, A_p) \\
            (b_q,B_q) \\
             \end{array}
   \right. \right] = c^{-\nu} \chi (-\nu),
\end{equation}
where
 \begin{equation}
\chi(s) = \frac{\prod_{j=1}^m \Gamma(b_j - B_j s) \prod_{j=1}^n \Gamma(1- a_j + A_j s)}{\prod_{j=m+1}^q \Gamma(1- b_j + B_j s) \prod_{j=n+1}^p \Gamma(a_j - A_j s)} .
\end{equation}
In what follows, we shall study the long-time behavior of the fractional moments for the cases where the medium growth is respectively dictated by a power law and by an exponential law.

\subsubsection{Power-law expansion}
\label{secPowerLawLevy}

In this case we assume that the uniform expansion is described by the power law
\begin{equation}
a(t)=\left(\frac{t+t_0}{t_0}\right)^{\gamma},
\label{a(t)PotExp}
\end{equation} where $t$ denotes the time elapsed since the initial time $t_0$.  In Cosmology, this power law is pertinent when the universe expansion
is dominated by matter $(\gamma=2/3)$ or by radiation $( \gamma=1/2 )$ \cite{Peacock1999, Ryden2003}.
We also note that negative values of $\gamma$ are also allowed and correspond to the situation of a contracting medium.

From Eqs.~\eqref{ConformalTime} and \eqref{a(t)PotExp} one finds the associated diffusion conformal time:
\begin{subnumcases}{\label{powerlawconftime} T(t)=}
\label{powerlawconftimea} \dfrac{t_0}{\gamma \mu - 1} \left[ 1 - \left( \dfrac{t+t_0}{t_0} \right)^{1 - \gamma \mu}  \right]  &   if \;\; $\gamma \mu \neq 1$, \\[2mm]
 t_0 \log \left( \dfrac{t+t_0}{t_0} \right)  &   if \;\; $\gamma \mu = 1 $.
\end{subnumcases}
This gives the following asymptotic long-time behavior:
\begin{subnumcases}
{T(t) \sim}
\dfrac{t_0}{\gamma \mu - 1} &   {if}   \;\;\;  $\gamma \mu > 1$,
 \\[2mm]
 t_0 \log \left( \dfrac{t+t_0}{t_0} \right) &   {if}   \;\;\;  $\gamma \mu = 1$,
 \\[2mm]
 \frac{t_0}{1 - \gamma \mu}  \left( \dfrac{t}{t_0} \right)^{1 - \gamma \mu}  &   if \;\;\; $\gamma \mu < 1$.
\end{subnumcases}
Insertion of the above expressions into Eq.~(\ref{ComovingNormalLevyFligthFractionalMoment}) gives
\begin{subnumcases}
{
\langle |x|^{\nu} (t) \rangle ^{\mu/\nu}
 \sim
 }
 C_{\nu\mu}\,\mathfrak{D}_{1,\mu}\, \dfrac{t_0}{\gamma \mu - 1}   &   {if}  \;\;\; $\gamma \mu > 1$,
\\[2mm]
 C_{\nu\mu}\,  \mathfrak{D}_{1,\mu}\, t_0\, \log \left( \dfrac{t+t_0}{t_0} \right)  &   {if}  \;\;\; $\gamma \mu = 1$,
\\[2mm]
 C_{\nu\mu}\,  \mathfrak{D}_{1,\mu} \, \dfrac{t_0^{\gamma\mu}}{1- \gamma \mu} \, t^{1 - \gamma \mu}  &   {if}  \;\;\; $\gamma \mu < 1$
\end{subnumcases}
in the limit $t\to\infty$.

Thus, any fractional moment with $0 < \nu < \mu$ displays a power law growth when $\gamma \mu < 1$, and a logarithmic power law growth when $\gamma \mu = 1$. However, it tends to a constant value when $\gamma \mu > 1$. This reflects the fact that the comoving propagator tends to a stationary profile $W_{st}(x)\equiv W(x,T(t\to \infty))$ in the long time limit. In turn, this is a direct consequence of the fact that $T$ saturates to a finite value if the expansion is sufficiently fast. In this case, the motion of the particle is dominated by the Hubble flux associated with the medium expansion rather than by the spread due to diffusion.

Finally, the behavior of the positional moment in the physical coordinate is straightforwardly obtained from the exact relation
\begin{equation}
\langle |y|^\nu (t) \rangle =a(t)^\nu \langle |x|^\nu (t) \rangle,
\label{fracmomcon}
\end{equation}
implying that $\langle |y|^\nu (t) \rangle$ grows as $t^{\gamma\nu}$ when  $\gamma \mu \ge 1$ (with a logarithmic correction in the marginal case $\gamma \mu=1$) and as $t^{{\nu}/{\mu}} $ when $\gamma \mu < 1$. In the former case, the time growth depends on the expansion exponent $\gamma$, but not on the L\'evy flight exponent $\mu$. As mentioned above, this means that the spreading of the particles is mainly driven by the expansion of the medium (i.e., by the Hubble flux), whereas the spreading due to the L\'evy diffusion process is negligible.  For the case with $\gamma \mu < 1$, the behavior is just the opposite: the growth of $\langle |y|^\nu (t) \rangle$ depends on $\mu$, but not on $\gamma$, which means that the spreading of the particles throughout  the medium is mainly due to diffusion. Finally, we note that the above results remain valid for the case of a contracting medium ($\gamma<0$).

\subsubsection{Exponential expansion}

Next, let us turn to the case where the scale function displays exponential behavior, i.e.,
\begin{equation}
a(t) = \exp\left( H t \right),
\label{a(t)ExpExp}
\end{equation}
with the ``Hubble constant" $H=\dot{a}(t)/a(t)$. In Cosmology, the above exponential law describes the regime where the universe expansion is driven by dark energy \cite{Peacock1999, Ryden2003}.

From Eq.~\eqref{ConformalTime}, the  conformal time  in this case is
 \begin{equation}
 T(t) = \frac{1- \exp(-H \mu t)}{H \mu}.
 \label{Texp}
 \end{equation}
This result is valid regardless of the sign of $H$. The case $H>0$ corresponds to an expanding medium (the situation we shall mainly focus on), whereas the case $H<0$ describes a contracting medium. In the former case, the conformal time eventually reaches a finite value, $T(t\to\infty)=(H \mu)^{-1} $, and, consequently,  a stationary comoving propagator settles. Insertion of this final value for $T(t)$  into Eq.~\eqref{ComovingNormalLevyFligthFractionalMoment} yields the long-time behavior of the fractional comoving $\nu$th moment:
 \begin{equation}
\langle |x|^{\nu} \rangle^{\mu/\nu} \sim  \frac{C_{\nu \mu}\,\mathfrak{D}_{1,\mu}}{H \mu}  .
\end{equation}
In contrast, for $H<0$ the fractional moments display exponential growth at long times, as is readily seen from Eqs.~\eqref{ComovingNormalLevyFligthFractionalMoment} and \eqref{Texp}.

Turning now to the analysis of the fractional moments $ \langle |y|^{\nu} (t) \rangle$ in physical space, one sees that the behavior is exactly the opposite, that is, for $H>0$ unbounded exponential growth is observed, whereas for $H<0$  the $\nu$th fractional moment attains a long-time asymptotic value given by the equation
\begin{equation}
\langle |y|^{\nu} \rangle^{\mu/\nu} \sim -\frac{C_{\nu \mu}\,\mathfrak{D}_{1,\mu}}{H \mu}.
\end{equation}
Thus, a stationary propagator in physical space is also attained in this case.  This point will be further discussed in Sec.~\ref{secOnsetStatProp}.

\subsection{Diffusion equation for L\'evy flights in physical coordinates}

In order to rewrite Eq.~\eqref{ComovingNormalLevyFligthFDE} in physical coordinates, we follow a procedure similar to the one used in Ref. \cite{Yuste2016}. The pdf $P(y,t)$ can be written as  $P(y,t)=P_x(y/a(t),t)$ in terms of the auxiliary function $P_x(x,t)=a(t) W(x,t)$. Writing $W(x,t)$ as  $P_x(x,t)/a(t)$ in Eq.~\eqref{ComovingNormalLevyFligthFDE} and using the relation

\begin{equation}
\frac{\partial P_x (x,t)}{\partial t} = \frac{\partial P (y,t)}{\partial t} + \frac{\dot{a}(t)}{a(t)} y  \frac{\partial P (y,t)}{\partial y}
\label{TimePartialPx}
\end{equation}
one finds

 \begin{equation}
\frac{\partial P}{\partial t} = - \frac{\dot{a}(t)}{a(t)} \frac{\partial}{\partial y} \left[ y P(y,t) \right] + \mathfrak{D}_{1,\mu} \nabla_y^\mu P(y,t).
\end{equation}
The only difference between the above equation and the diffusion equation given in Ref. \cite{Yuste2016} is that here $\nabla_y^\mu$ replaces the usual Laplacian operator in the second term of the rhs.  In contrast, the first term was already present in the diffusion equation of Ref. \cite{Yuste2016}, and describes the deterministic drift and the dilution effect arising from the medium expansion.

\subsection{Onset of stationary propagators}
\label{secOnsetStatProp}
As already anticipated, for sufficiently fast expansions we obtain long-time stationary propagators in comoving coordinates. This is consistent with one's intuitive expectation, since any time-independent  function of the position of a particle in the comoving space stems from its lack of intrinsic movement in the physical space. Thus, the onset of stationary propagators in the comoving space for sufficiently long times tells us that the expansion of the medium is so fast that the intrinsic motion of the particles due to diffusion becomes negligible. On the other hand, it is interesting to note that in the case of an exponential contraction, a stationary propagator is also observed in physical space.

For a power-law expansion with $\gamma \mu > 1$, a simple substitution of the asymptotic value of the diffusion conformal time $ \lim_{t \to \infty} T(t) = t_0 / (\gamma \mu -1)$ in Eq.~\eqref{solFExT} yields the stationary propagator in comoving coordinates, namely,
 \begin{equation}
 W_{st} (x) = \mathsf{L}_{\mu} \left(  x;  \frac{\mathfrak{D}_{1,\mu} t_0 }{\gamma \mu -1}  \right) .
\end{equation}

On the other hand, for an exponential expansion with $H>0$, one has  $T(\infty) = 1/ (H \mu)$ and the comoving stationary propagator has the form
\begin{equation}
W_{st} (x) = \mathsf{L}_{\mu} \left(  x;  \frac{\mathfrak{D}_{1,\mu} }{H \mu} \,\right) .
\end{equation}

In the case of exponential contraction ($H<0$), the diffusion conformal time behaves as $T(t) \sim [1- \exp(-H \mu t)] / (H \mu)$ and there is no stationary comoving propagator. In contrast, a stationary propagator in physical space $P_{st} (y)$ is observed in the long time limit. Indeed, for an arbitrary time $t$, and using Eq.~\eqref{solFExT}, one has
 \begin{equation}
P (y,t) = \frac{1}{a(t)} W \left[\frac{y}{a(t)}, t\right]= \frac{1}{a(t)}\, \mathsf{L}_{\mu} \left[ \frac{y}{a(t)};  \mathfrak{D}_{1,\mu} T \right] .
\label{transientprop0}
\end{equation}
Taking into account the scaling property \eqref{LevyScaling},
one finds that Eq.~\eqref{transientprop0} can be rewritten as
\begin{equation}
P (y,t) =\mathsf{L}_{\mu} \left[y; a^\mu(t) \mathfrak{D}_{1,\mu} T\right].
\label{transientprop}
\end{equation}
At long times, one has
$
\lim_{t \to \infty} \left[  T(t) a^\mu(t) \right] =  -1/(H \mu)$,
implying that the stationary propagator takes the form
\begin{equation}
P_{st} (y) = \mathsf{L}_{\mu} \left( y;  -\frac{\mathfrak{D}_{1,\mu}}{H\mu}  \right).
\label{Psty}
\end{equation}
\begin{figure}[t]
\includegraphics[width=0.48\textwidth]{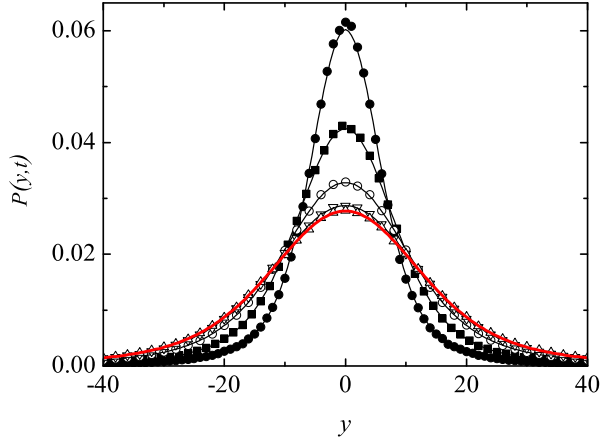}
\caption{Time evolution of the physical propagator for $H=-10^{-2}$ and L\'evy exponent $\mu=3/2$. Solid thin lines represent the analytical propagator $P(y,t)$ for $t=25, 50, 100, 200, 500$ (from top to bottom at $y=0$). The corresponding symbols (solid circle, solid square, open circle, down triangle and up triangle) represent simulation results obtained with $10^6$ realizations.  The thick line represents the asymptotic theoretical stationary propagator given by Eq.~\eqref{Psty} and overlaps with the thin line corresponding to the analytical result for $t=500$.  In the simulations we used  $\lambda(y)=\mathsf{L}_{3/2}\left(y;\sigma^{3/2}=1/2\right)$ and a waiting time pdf $\varphi(t)$ given by the exponential pdf  ~\eqref{expphi}  with $\tau=1$. According to Eq.~\eqref{diffconst}, this parameter choice implies $\mathfrak{D}_{1,3/2}=1/2$.}
\label{FigPHm001}
\end{figure}
Note that the replacement $H\to -H$ in the stationary propagator for the case of exponential contraction allows one to recover the comoving propagator for the case of an exponential expansion.
Equation~\eqref{Psty} is a remarkable result, since a stationary state of the diffusion process arises despite the absence of any external force.  In the present case, the onset of a stationary propagator is a consequence of the competition between the spreading of particles due to diffusion and the medium contraction, which results in a kind of balance between both effects, whereby the contracting Hubble flux plays the role of a fictitious external force that eventually drives the system to a stationary state.

In Fig.~\ref{FigPHm001}  we show the convergence of $P(y,t)$ (as given by Eq.~\eqref{transientprop} and simulation results) to the stationary propagator \eqref{Psty}. Note that the superdiffusive particle spreading is increasingly counterbalanced by the medium contraction and significantly slowed down at long times. Finally the transient propagator becomes indistinguishable from the stationary one.

\subsection{Width of the propagator}
\label{secwidthPropagator}

Typically, the width of the propagator is characterized by its second moment, but in the case of L\'evy flights in an infinite medium the latter does not exist. Hence, one must resort to an alternative definition of the typical width characterizing the diffusive spread of particles. Here, we choose the time-dependent quantity $\Delta_x $, which defines a symmetric interval [-$\Delta_x$/2,$\Delta_x$/2] about the origin so that the probability to find the walker within this interval is $1/2$. Therefore, $\Delta_x$ is defined via the equation
 \begin{equation}
\int_{{- \Delta_x}/{2}}^{{\Delta_x}/{2}} \, W(x,t) dx = \frac{1}{2},
\end{equation}
$W(x,t)$ being the propagator in comoving coordinates. This equation implies
\begin{equation}
 \Delta_x =w_{\mu} \left[ \mathfrak{D}_{1,\mu} T(t) \right]^{{1}/{\mu}}.
 \label{TimeDepWidth}
 \end{equation}
 In the special cases $\mu=1$ and $\mu=2$, the proportionality constant $w_{\mu}$ can be exactly computed and one has $w_1 = 2$ and $w_2 = 4 \, \mbox{erf}^{-1}(1/2)$. In other cases, the value of this constant can be found numerically. In physical space, the corresponding width  $\Delta_y$ is simply $a(t)\Delta_x$.

In a static medium, the width of the L\'evy propagator can be shown to grow as $t^{1/\mu}$ \cite{Jespersen1999, Metzler2000}. In contrast, for a power-law expansion, insertion of Eq.~\eqref{powerlawconftime} into Eq.~\eqref{TimeDepWidth} shows that several cases can once again be distinguished as a result of the interplay between diffusive spreading and the medium expansion. When $\gamma{\mu} > {1}$, a stationary propagator in the comoving space arises. For sufficiently slow expansions ($\gamma {\mu}< {1}$), the medium growth can be regarded as a weak perturbation with respect to the case of diffusive transport in a static medium, and the $t^{1/\mu}$-dependence of the propagator width is preserved. In contrast, a fast expansion with $\gamma {\mu}> {1}$ makes the contribution of the intrinsic diffusion process to the physical displacement irrelevant, and the growth of $\Delta_y$ as $t^{\gamma}$ at long times is basically due to the medium expansion.

Finally, let us discuss the behavior of the propagator width in the case of an exponential expansion and of an exponential contraction. In the former case, diffusion is a very minor perturbation, and the propagator in physical space displays an exponential blow-up induced by the medium expansion.  In the comoving frame, $\Delta_x$ goes to a constant value at long times. In the opposite case of a contracting medium, one has exponential widening of the propagator in the comoving space. As already mentioned above, the competition between diffusive spreading and the medium contraction results in the onset of a stationary physical propagator at long times.

\subsection{L\'{e}vy horizon}

We now proceed to discuss how the expansion of the medium affects the mixing properties of L\'{e}vy diffusion processes. The computation of the probability that a random walker has reached a specific spatial domain after a given time, or of the probability that she ever reaches that domain, is directly related to the computation of the encounter probability for two diffusing particles starting at different positions. All these probabilities underlie basic problems in reaction-diffusion kinetics, and it is clear that their values will depend on whether the medium expands or not. In fact, if the expansion is fast enough, the probability that a walker ever reaches a certain domain is vanishingly small, as is the probability that two walkers ever meet. In what follows, we shall address this question by establishing a quantitative connection between the expansion rate and the reach of a L\'{e}vy diffusion process.

To this end, we shall first introduce the preliminary notions of ``diffusive pulse'' and ``L\'{e}vy horizon''. In the same way as a light pulse is defined as the signal carried by photons starting from the same initial position, a diffusive pulse is defined here as the signal transmitted by walkers starting from the same initial position.  Clearly, the reach of a diffusive pulse can then be quantified by the width $\Delta_x$ of the comoving propagator  (see section \ref{secwidthPropagator}). At this stage, we note that, unlike the photons of a light pulse, some of the walkers of a diffusive pulse (in fact, one half on average) travel beyond the typical distance $\Delta_x$; therefore, the notion of ``reach'' in this context is to be understood in a statistical sense only.

In what follows, the quantity  $\Delta_x/2$  will be termed ``L\'{e}vy horizon'' and denoted by   $x_\text{Lh}(t)$. Besides, the $t\to\infty$ limit of $x_\text{Lh}(t)$ will be termed ``L\'{e}vy event horizon'' and denoted by  $x_\text{Leh}$ hereinafter, i.e., $x_\text{Leh}\equiv x_\text{Lh}(t\to \infty)$.  In other words, $x_\text{Leh}$ is the typical comoving distance traveled by a diffusive L\'{e}vy pulse during an infinite time. This means that regions separated by a distance significantly larger than the L\'{e}vy  horizon $x_\text{Lh}(t)$ cannot be connected by a L\'{e}vy diffusion process up to time $t$, in the sense that the diffusive particles initially found in each of these distant regions cannot mingle and are thus prevented from taking part in any kind of diffusion-controlled reaction up to that  time $t$. In other words, for a sufficiently fast medium expansion, if the initial separation between two diffusive pulses exceeds the  threshold value $x_\text{Lh}(t)$, mixing effects become negligible up to time $t$. Note that we have coined the term ``L\'{e}vy event horizon'' by analogy with the definition of ``event horizon'' in Cosmology, which is the largest comoving distance that a light pulse emitted at the initial time $t_0$ can ever reach (an observer located at a distance from the light source larger than the event horizon is not able to measure any signal). For this reason, the quantities $x_\text{Lh}(t)$ and $x_\text{Leh}$  are useful quantifiers that allow one to predict mixing properties of L\'{e}vy processes in expanding media.

A particular example is given by two delta pulses initially located at positions $\pm x_0$. The solution of Eq.~(\ref{ComovingNormalLevyFligthFDE}) satisfying this double-peaked initial condition $W(x,0) = [\delta(x-x_0) + \delta(x+x_0)]/2$ is given by a simple linear superposition, i.e., $W_2 (x,t) =  [W(x-x_0,t) + W(x+x_0,t) ]/2$, where $W(x,t)$ is the solution of Eq.~(\ref{ComovingNormalLevyFligthPropagator}) for a delta-peaked initial condition $\delta(x)$.

\begin{figure}[t]
{\includegraphics[width=0.48\textwidth]{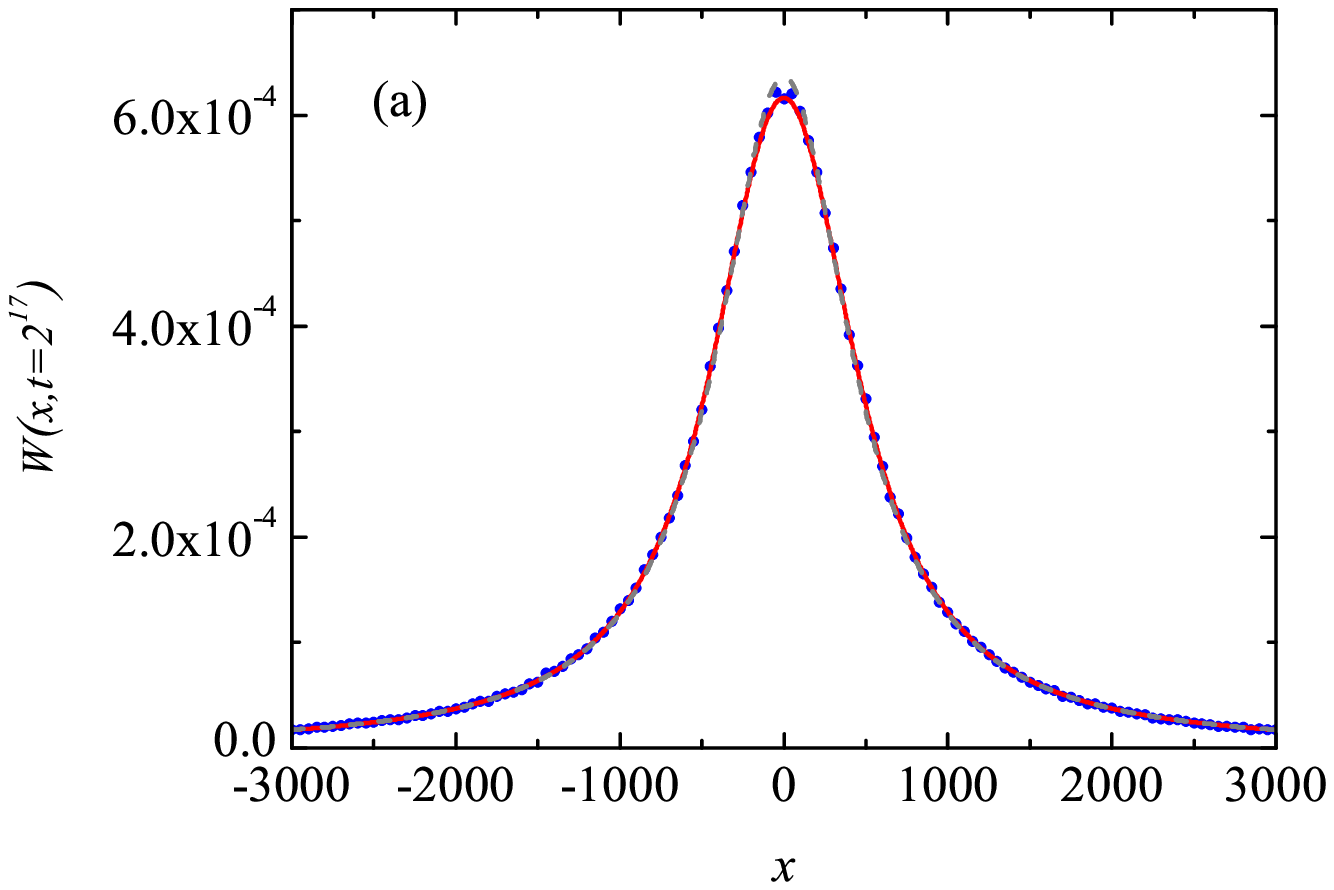}} \hfill
{\includegraphics[width=0.48\textwidth]{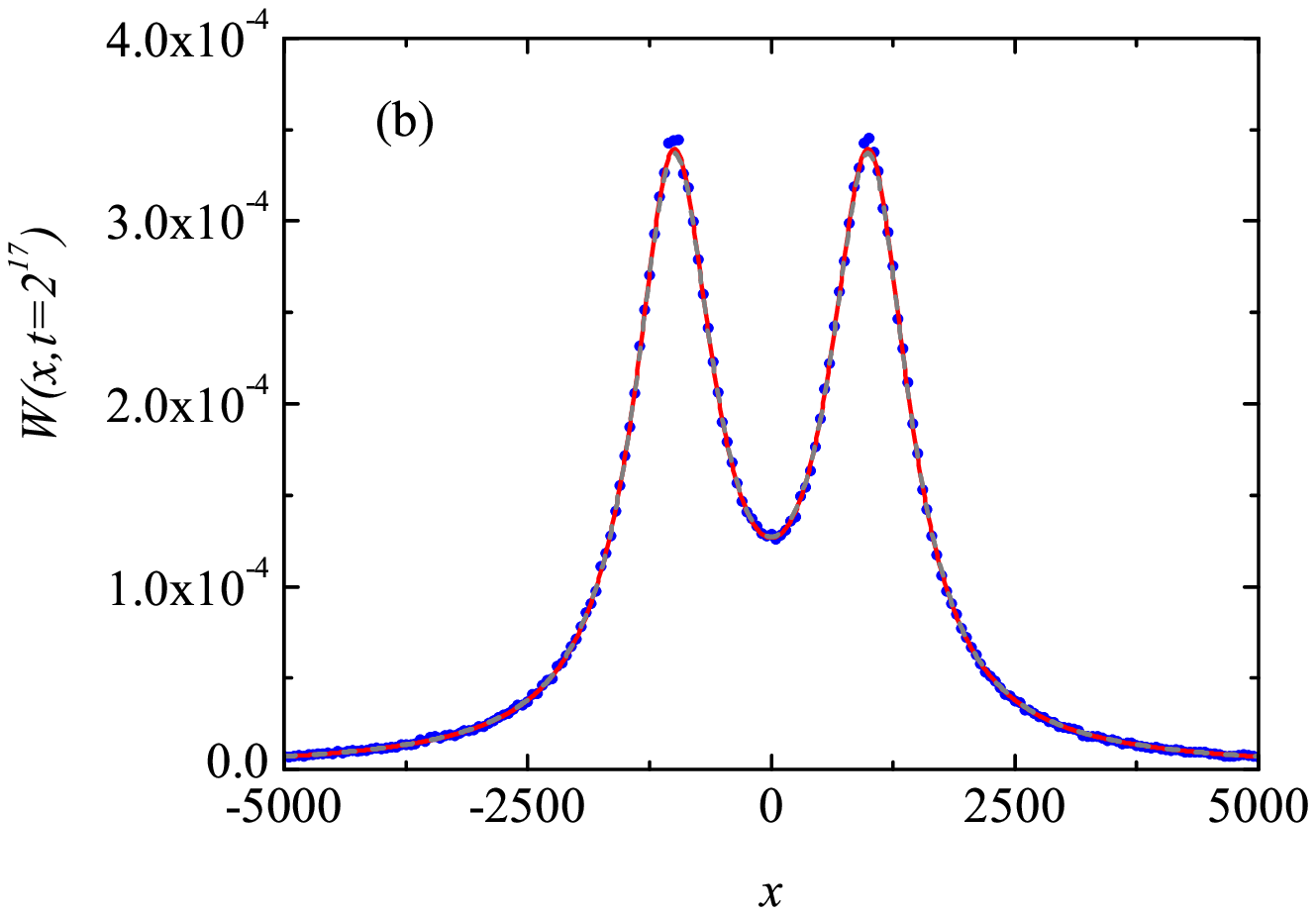}}\hfill
{\includegraphics[width=0.48\textwidth]{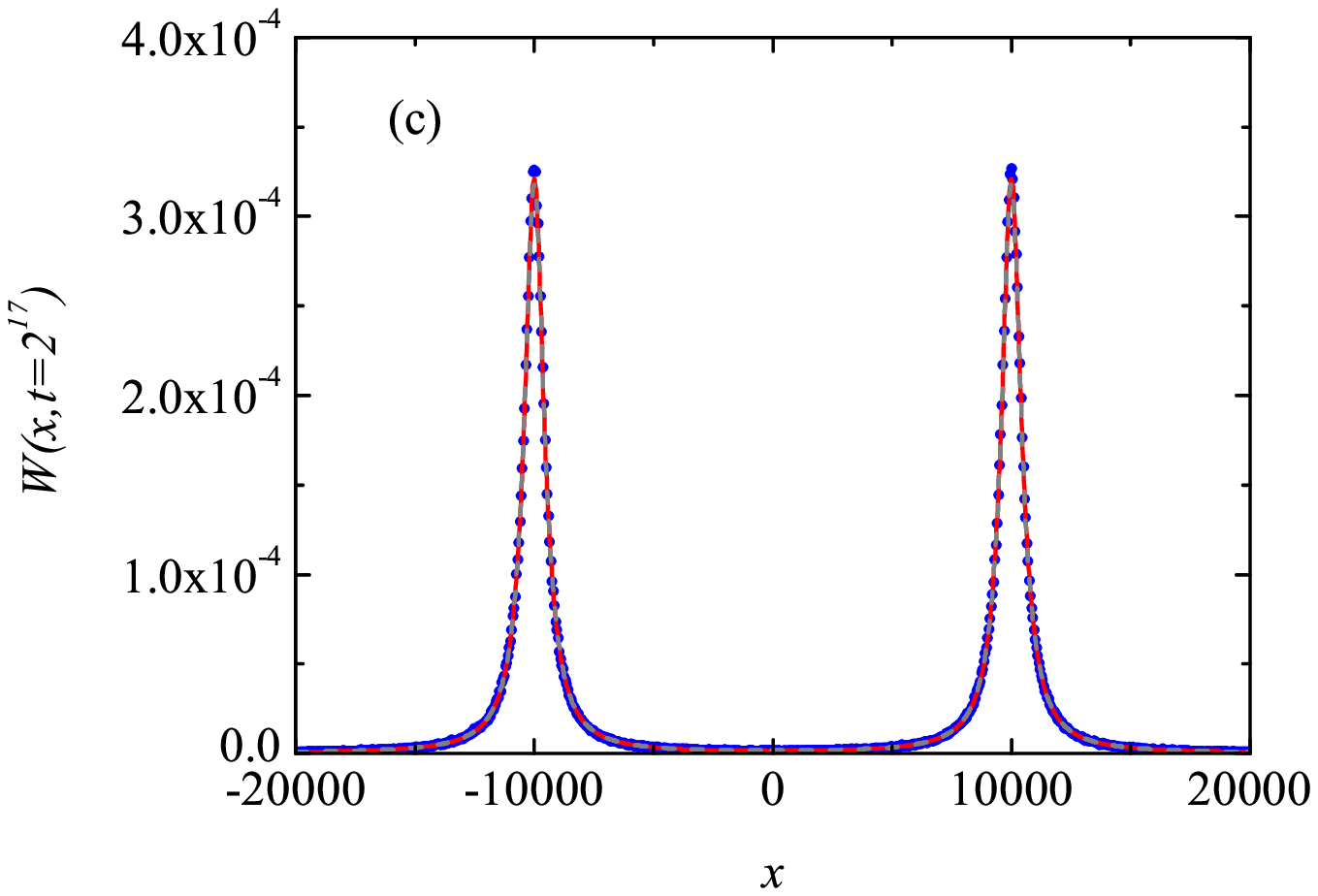}}\hfill
\caption{
Pdf corresponding to  two L\'evy pulses with $\mu=1$ (Cauchy pulses)  at $t=2^{17}$  in the case of a power-law expansion with $\gamma=2$,  $t_0=10^3$ and $\mathfrak{D}_{1,1}=1/2$. This implies $x_\text{Lh}(2^{17})=496.2$  and  $x_\text{Leh}=500$. The pulses depart from $x=\pm x_0$ with (a) $x_0=10^2$ (merged pulses), (b) $x_0=10^3$ (mutually influenced pulses), and (c) $x_0=10^4$ (clearly separated pulses).
Solid lines represent the analytical propagator $W_2\left(x,t=2^{17}\right)$, dashed lines depict the asymptotic stationary state and dots  represent simulation results ($10^6$ realizations) with $\lambda(y)$ given by  Eq.~\eqref{Cauchypdf} with $\sigma=1/2$ and $\varphi(t)$ given by  Eq.~\eqref{expphi}  with $\tau=1$.   The solid and dashed lines are almost indistinguishable from one another because the chosen time $t=2^{17}$ is so large that the propagator $W\left(x,t=2^{17}\right)$  is already very close to the asymptotic one.
 }
\label{Fig:LevyPulsesComovingPropagators}
\end{figure}

Let us discuss the behavior of the two pulses depending on the prescribed values of $x_0$ and $x_\text{Leh}$.  When $x_\text{Leh} \ll x_0$, the two pulse pdf converges to a long-time profile $W_2 (x,\infty)$ characterized by two clearly distinct peaks. In contrast, when $x_\text{Leh} \gg x_0$, both pulses merge and the final profile approaches a unimodal pdf which closely follows the long-time asymptotic form of the free solution for the initial condition $\delta (x)$ in the limit $x_0 \to 0$. Finally, if the L\'evy event horizon is of the order of $x_0$, a kind of intermediate behavior is observed. While two distinct peaks are clearly distinguished in the stationary profile, it is also true that the probability to find the particle near the origin $x=0$ becomes non-negligible. This qualitative behavior is illustrated in Fig.~\ref{Fig:LevyPulsesComovingPropagators}, where the comoving propagator is represented for these three different situations. The simulation results provided in this figure have been generated by means of the exponential  waiting time pdf given by Eq.~\eqref{expphi}  with $\tau=1$ and the Cauchy  jump length pdf
\begin{equation}
\lambda(y)=\frac{\sigma}{\pi(\sigma^2+y^2)}
\label{Cauchypdf}
\end{equation}
with $\sigma=1/2$. According to Eq.~\eqref{diffconst}, this parameter choice corresponds to an anomalous diffusion constant $\mathfrak{D}_{1,1}=1/2$. This choice   of $\tau$ and $\sigma$ defines our  time and length units, respectively.
For these parameters [see Eq.~\eqref{TimeDepWidth}] the L\'evy horizon is just $x_\text{Lh}(t)=\Delta_x/2=T(t)/2$ where $T(t)$ is given by Eq.~\eqref{powerlawconftimea}. In particular, $x_\text{Lh}(2^{17})=496.2$  and $x_\text{Leh}=x_\text{Lh}(\infty)=500$.

Finally, we note that in the case of a power-law expansion finite L\'evy event horizons may only arise when $\gamma\mu > 1 $ (see section \ref{secwidthPropagator}). In this case,  the diffusion conformal time converges to a finite value $t_0 /(\gamma \mu -1)$ at long times. As for the case of an exponential expansion, a finite L\'evy event horizon may only arise when $H>0$.

In all other cases the L\'evy event horizon is infinite and significant diffusive mixing will eventually take place irrespective of the initial separation of both particles.

\section{The subdiffusive CTRW in an expanding medium}
\label{secIV}

Having discussed in detail the case of L\'evy flights on the basis of  Eq.~\eqref{ComovingNormalLevyFligthFDE}, we shall next focus on a different anomalous diffusion case, namely, on the subdiffusive CTRW, which is the oldest and most studied CTRW model. Specifically, we consider the case of a waiting time pdf $\varphi(\Delta t)$ displaying the power-law long-time decay described by Eq.~\eqref{longtailphi} and a symmetric distribution $\lambda(\Delta y)$ with finite variance,  such as the Gaussian pdf \eqref{gaussianLambda1}. In this case, we recall that Eq.~\eqref{bifractionaleq} takes the form

\begin{equation}
\frac{\partial W (x,t)}{\partial t} = \mathfrak{D}_{\alpha} \frac{1}{a^2(t)} ~ \frac{\partial^2 }{\partial x^2} \left[ {_0}\mathcal{D}_t^{1-\alpha} W(x,t) \right],
\label{AnomalousExpansiveDiffusionEquation2}
\end{equation}
where $\mathfrak{D}_{\alpha} \equiv \mathfrak{D}_{\alpha,2}=\sigma^2 / \tau^{\alpha} $. Formally, the above equation is a FDE with an effective time dependent diffusion coefficient, a type of problem that has already been considered in the literature \cite{Magdziarz2014,Hristov2017}.

\subsection{Moments}

As is typical for diffusion problems, the  comoving moments of the pdf $W(x,t)$ are connected to one another via a descending hierarchy of differential equations. In the present case one has the set of equations
\begin{equation}
\frac{d}{dt} \langle x^m (t) \rangle = m (m-1) \frac{\mathfrak{D}_{\alpha}}{a^2(t)}~{_0}\mathcal{D}^{1-\alpha}_t \langle x^{m-2} (t) \rangle, \qquad m=2,4,\ldots,
\label{mthComovingMoment}
\end{equation}
which is obtained by multiplying Eq.~\eqref{AnomalousExpansiveDiffusionEquation2} with $x^m$ and by subsequently integrating the resulting equation over the space.  Odd moments vanish due to the absence of a bias. For the second moment ($m=2$) one obtains
\begin{equation}
\frac{d}{dt} \langle x^2 (t) \rangle = \frac{\mathfrak{D}_{\alpha}}{a^2(t)} ~ {_0}\mathcal{D}_t^{1-\alpha} (2) =  \frac{2 \mathfrak{D}_{\alpha}}{a^2(t)} \frac{t^{\alpha - 1}}{\Gamma(\alpha)}.
\label{SecondOrderComovingMomentum}
\end{equation}
As done previously, we now proceed to discuss the behavior in the special cases of a power-law expansion and of an exponential expansion. In the former case we discuss the behavior of the second order moment, whereas in the latter case one can easily account for the long-time behavior of the full set of moments by taking advantage of the shift theorem for the Laplace transform.

\subsubsection{Power-law expansion}

\begin{figure}[t]
 {\includegraphics[width=0.48\textwidth]{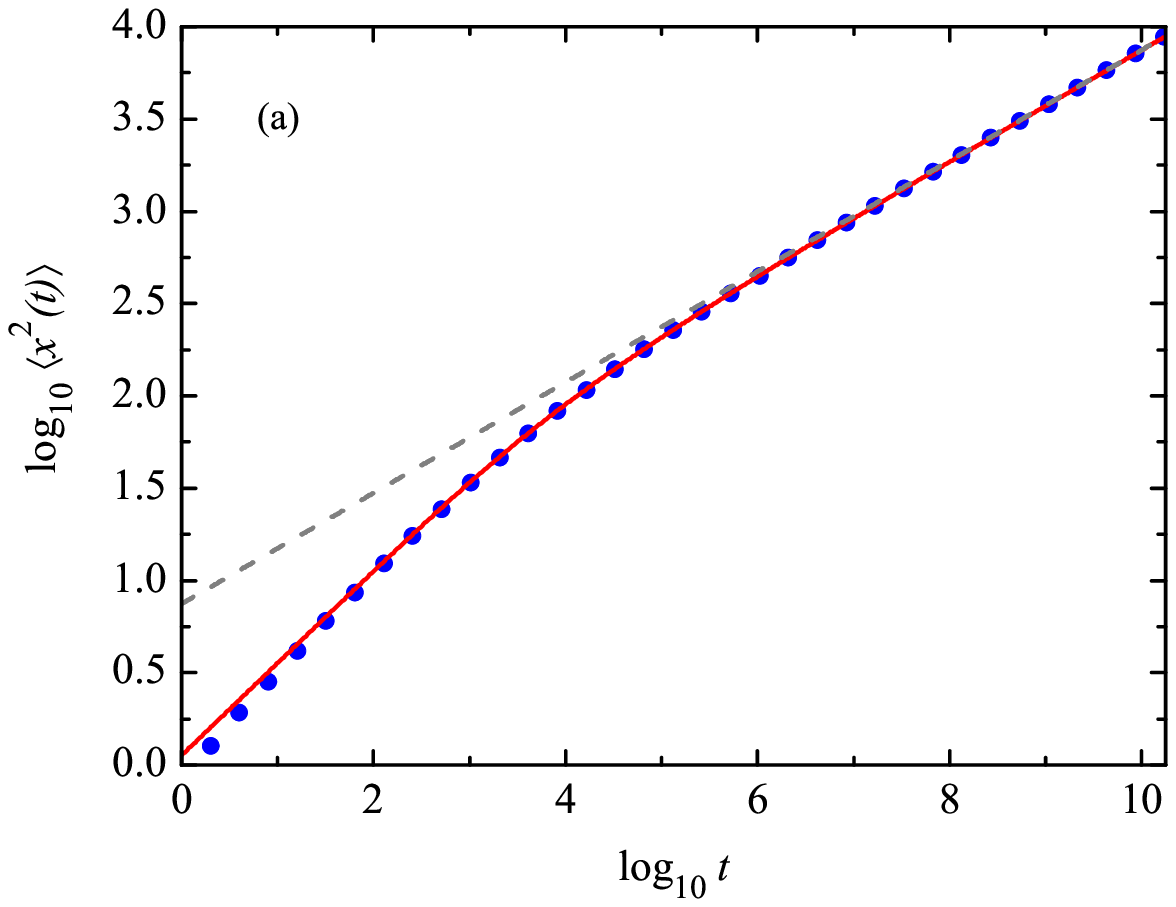}}\hfill
 {\includegraphics[width=0.48\textwidth]{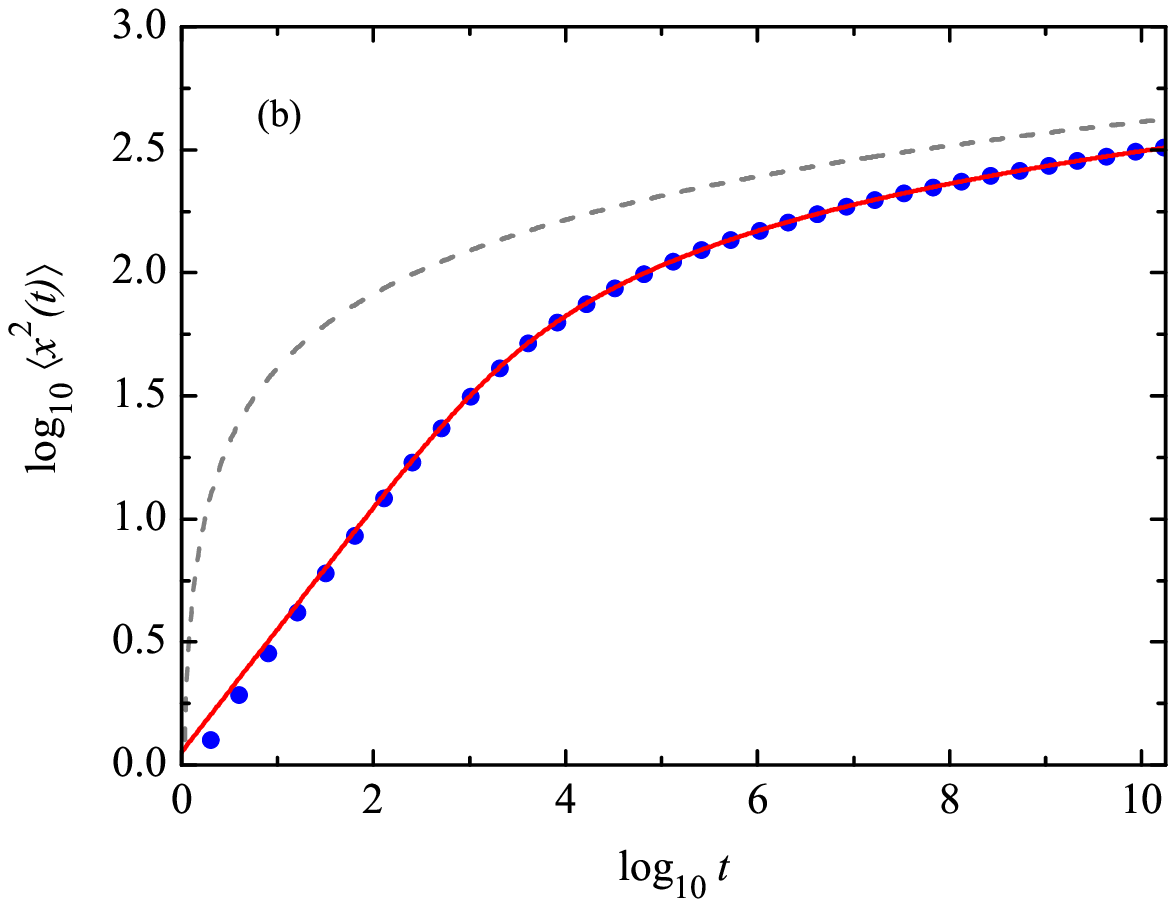}}\hfill
 {\includegraphics[width=0.48\textwidth]{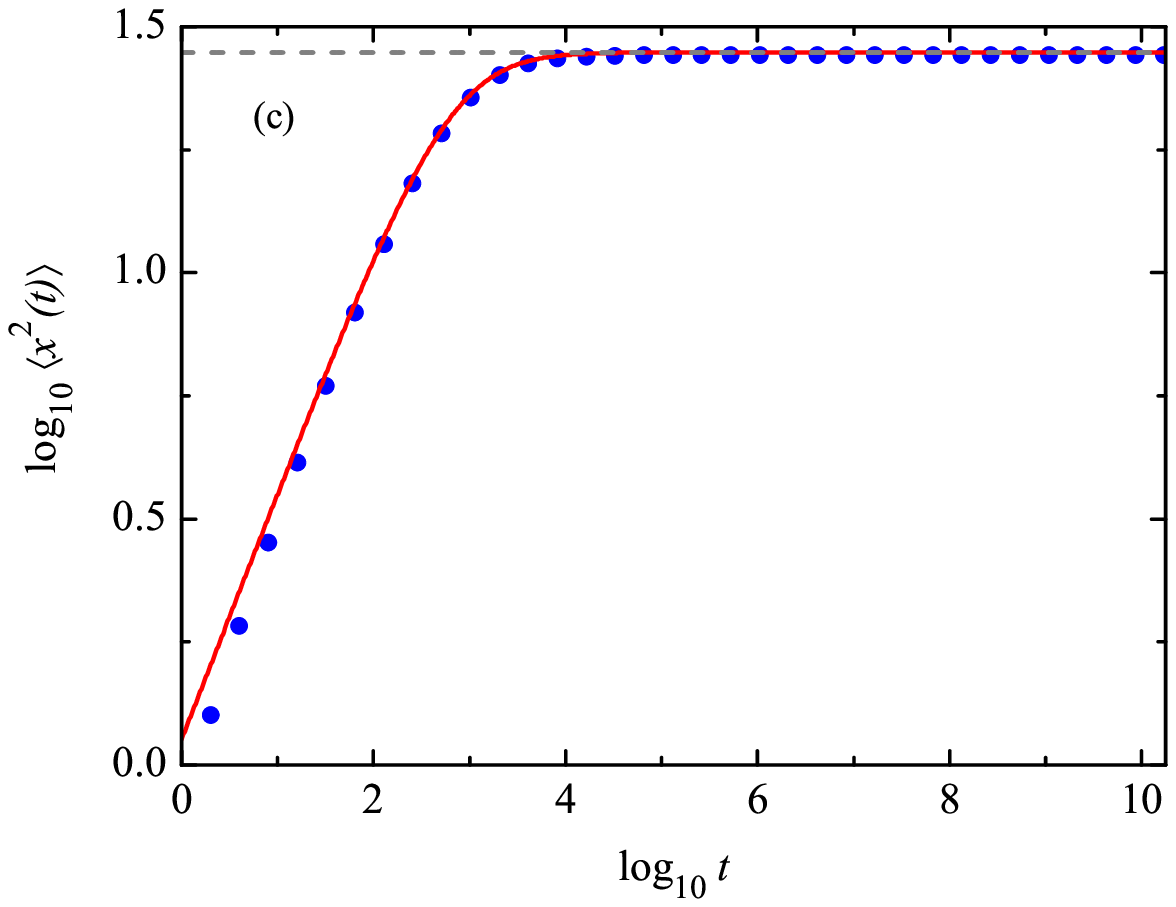}}\hfill
\caption{Double-logarithmic representation of the comoving second order moment for a subdiffusive CTRW with $\alpha=1/2$ in a medium subjected to a power-law expansion with $t_0 =10^3$ and
(a) $\gamma=1/10$ (diffusion-dominated regime),  (b) $\gamma=\alpha/2=1/4$ (marginal case), and  (c)  $\gamma=1$ (expansion dominated regime).  Solid lines represent analytical results obtained from Eq.~(\ref{SecondOrderComovingMomentumPotential}), whereas dashed lines represent the long-time asymptotics described by Eq.~\eqref{AsymptoticSecondOrderComovingMomentumPotentialExp}. The dots correspond to numerical simulations performed up to $t=2^{34}$ ($10^6$ realizations) with $\lambda(y)$ given by  Eq.~\eqref{gaussianLambda1} with $\sigma^2=1/2$ and $\varphi(t)$ given by Eq.~\eqref{paretoSimu}  with $\xi=1/\pi$.
}
\label{Fig:ComovingSecondMomentSubdiffusionPotentialExp}
\end{figure}

Inserting Eq.~(\ref{a(t)PotExp}) into Eq.~(\ref{SecondOrderComovingMomentum}) yields the differential equation
 \begin{equation}
\frac{d}{dt} \langle x^2 (t) \rangle =   \frac{2 \mathfrak{D}_{\alpha}t_0^{2\gamma} }{\Gamma(\alpha)} t^{\alpha - 1} (t+t_0)^{-2\gamma},
\label{SecondOrderComovingMomentumPotential}
\end{equation}
from which the time dependence of the second-order moment in the comoving space can be studied. Integration of the above equation yields

 \begin{equation}
\langle x^2 (t) \rangle = \frac{2 \mathfrak{D}_{\alpha}}{\alpha \Gamma(\alpha)} t^{\alpha}  {_{~ 2}}F_1 \left( \alpha,2\gamma;1+\alpha; \frac{-t}{t_0} \right)
\label{SecondOrderComovingMomentumPotentialExp},
\end{equation}
where $_2F_1 $ stands for the ordinary hypergeometric function. It is instructive to address the behavior in different time regimes by means of suitable approximations based on the asymptotic behavior of $_2F_1 $ \cite{Abramowitz1972}. For very long times ($t \gg t_0$) one obtains
\begin{subnumcases}
{\label{AsymptoticSecondOrderComovingMomentumPotentialExp}
\langle x^2 (t) \rangle \sim }
             \dfrac{2 \mathfrak{D}_{\alpha}t_0^{2\gamma} }{\Gamma(\alpha) (\alpha-2\gamma)} t^{\alpha-2\gamma} &   {if}  \;\; $\alpha > 2 \gamma$,
\\[3mm]
         \dfrac{2 \mathfrak{D}_{\alpha}t_0^{\alpha} }{\Gamma(\alpha)} \log(t) &  {if} \;\; $\alpha=2\gamma$,
\\[3mm]
 \dfrac{\Gamma(2\gamma-\alpha) 2 \mathfrak{D}_{\alpha}t_0^{-\alpha} }{\Gamma(2\gamma)} &  {if}  \;\; $\alpha<2\gamma$.
\end{subnumcases}
As shown in Fig.~\ref{Fig:ComovingSecondMomentSubdiffusionPotentialExp}, Eqs.~\eqref{SecondOrderComovingMomentumPotentialExp} and
\eqref{AsymptoticSecondOrderComovingMomentumPotentialExp}
are in excellent agreement with simulation results.
These simulation results have been generated by means of
the Gaussian  jump length pdf given by Eq.~\eqref{gaussianLambda1} with $\sigma^2=1/2$ and the Pareto waiting time pdf
\begin{equation}
\varphi(\Delta t) = \frac{\alpha}{\xi[1+(\Delta t/\xi)]^{1+\alpha}}
\label{paretoSimu}
\end{equation}
with  $\alpha=1/2$ and $\xi=1/\pi$. This choice leads to Eq.~\eqref{varphiAsin} with $\tau=1$.  According to Eq.~\eqref{diffconst}, these parameters   corresponds to the anomalous diffusion constant $\mathfrak{D}_{\alpha}=1/2$. Of course, this choice   of $\tau$ and $\sigma$ defines our  time and length units, respectively.

For the long-time behavior of the second order moment $\langle y^2 (t) \rangle= a^2(t) \langle x^2(t) \rangle$ in physical space, this implies $\langle y^2 (t) \rangle \propto t^\alpha$ for $2 \gamma < \alpha$, $\langle y^2 (t) \rangle \propto t^\alpha \log(t) $ for $\alpha = 2 \gamma$, and $\langle y^2 (t) \rangle \propto t^{2\gamma}$ for $2 \gamma >\alpha$.  The interpretation of this result is similar to the one given at the end of Sec.~\ref{secPowerLawLevy} for L\'evy flights.
One sees that $\langle y^2 (t) \rangle$ does not depend on the expansion exponent $\gamma$ when $2\gamma <\alpha$.  In fact, the qualitative behavior of the mean square displacement in physical space is similar to the case of a static medium, but one has a modified diffusion coefficient $\mathfrak{D}_{\alpha}^{\text{eff}} = \mathfrak{D}_{\alpha} {\alpha}/{(\alpha - 2\gamma)}$.
In contrast, for $2\gamma>\alpha$, the typical distance traveled by the walker is $\sqrt{ \langle y^2(t) \rangle } \sim t^{\gamma}$, which essentially means that its motion is governed by the medium expansion, the spread due to the subdiffusion process being negligible.

\subsubsection{Exponential expansion}

In this case,  Eq.~\eqref{SecondOrderComovingMomentum} becomes
\begin{equation}
\frac{d}{dt} \langle x^2 (t) \rangle = \frac{2 \mathfrak{D}_{\alpha}}{ \Gamma(\alpha)} t^{\alpha-1}  \exp \left( -2H t \right).
\label{Eq2ndmoment}
\end{equation}
The solution of Eq.~\eqref{Eq2ndmoment} is
\begin{equation}
\langle x^2 (t) \rangle = {2 \mathfrak{D}_{\alpha}} (2H)^{- \alpha}   \, \frac{\gamma(\alpha,2Ht)}{ \Gamma(\alpha)},
\label{SecondOrderComovingMomentumExpExp}
\end{equation}
where
\begin{equation}
\gamma (\mu,z)=\int_0^{z} t^{\mu-1} \exp(-t) dt
\end{equation} denotes the lower incomplete Gamma function.

It is worth noting that for $H>0$ the comoving variance tends to a finite value (this behavior also extends to even higher order moments). Specifically, $\lim_{t \to \infty} \langle x^2 (t) \rangle = 2 \mathfrak{D}_{\alpha} (2H)^{- \alpha} $. In contrast, it displays unlimited growth for $H \leq 0$ as expected (see also Sec.~\ref{secBigCrunch}).

Next, let us consider the behavior of higher-order moments.
The solution of Eqs.~\eqref{mthComovingMoment} becomes rapidly cumbersome in the real time domain. Fortunately, the calculation is greatly simplified in Laplace space by virtue of the shift theorem for the Laplace transform,
$\mathcal{L} [\exp(-2Ht) f(t)] = \tilde{f} (s+2H)$
(to simplify our notation, tildes denoting the Laplace transform will be omitted hereinafter).
Thus, in Laplace space we find that the hierarchy of moments (\ref{mthComovingMoment}) is converted into the following recurrence relation:
\begin{subequations}
\begin{equation}
\langle x^m (s) \rangle  = m (m-1) \mathfrak{D}_{\alpha} s^{-1} (s+2H)^{1-\alpha} \langle x^{m-2} (s+2H) \rangle , \quad m=4,6,8...
\end{equation}
\label{mthComovingMomentLaplace}
and
\begin{equation}
\langle x^2 (s) \rangle  =2 \mathfrak{D}_{\alpha} (s+2H)^{-\alpha} s^{-1}.
\end{equation}
\end{subequations}
Moreover, since in Laplace space physical moments are connected with the comoving moments via the equation

\begin{equation}
\langle y^m (s) \rangle = \mathcal{L} \left[ \exp(mHt) \langle x^m (t) \rangle \right] = \langle x^m (s-mH) \rangle,
\end{equation}
one straightforwardly finds from Eqs. (\ref{mthComovingMomentLaplace}) the following hierarchical relation:
\begin{subequations}
\begin{equation}
\langle y^m (s) \rangle  = m (m-1) \mathfrak{D}_{\alpha} (s-mH)^{-1} (s+2H-mH)^{1-\alpha} \langle y^{m-2} (s) \rangle , \quad m=4,6,8...
\end{equation}
\label{mthPhysicalMomentLaplace}
and
\begin{equation}
\langle y^2 (s) \rangle =  \langle x^2 (s-2H) \rangle = 2 \mathfrak{D}_{\alpha} (s-2H)^{-1} s^{-\alpha}.
\end{equation}
\end{subequations}
We shall revisit the above hierarchy of equations at a later stage to extract relevant information for the asymptotic long-time behavior of the physical moments.

As already anticipated, when $H>0$ all the even comoving moments are non-vanishing and converge to finite values $\langle x^m  (t\to\infty)\rangle$ in the limit $t\to\infty$. The latter are readily obtained from the relation $\lim_{t\to \infty}  \langle x^m (t) \rangle =\lim_{s\to 0} s\langle x^m (s)\rangle $. For the fourth-order moment, say, one has

\begin{equation}
\langle x^4 (s) \rangle = 24 \mathfrak{D}_{\alpha}^2  (s+2H)^{\alpha} (s+4H)^{\alpha} s^{-1} =  24 \mathfrak{D}_{\alpha}^2  (s^2+6Hs+8H^2)^{\alpha} s^{-1}.
\end{equation}
Hence, $ \langle x^4 (t) \rangle \sim 3 \mathfrak{D}_{\alpha}^2 8^{1-\alpha} H^{-2\alpha} $ when $t \to \infty$.

In fact, from the solution of the recurrence relation, it is possible to obtain a general expression for the asymptotic value of the comoving moments:
\begin{equation}
\label{xinftMoments}
\langle x^{2m} (t \to \infty) \rangle = 2m! (2m!!)^{-\alpha} \mathfrak{D}_{\alpha}^{m} H^{-m \alpha}, \quad m=1,2,3,\ldots.
\end{equation}
In turn, the Fourier-transformed propagator may be expanded in terms of the moments as follows:
\begin{equation}
\widehat{W}(k,t) = \int_{-\infty}^{\infty} \exp(-ikx) W(x,t) dx = \sum_{m=0}^{\infty} \frac{(-i k)^m}{m!} \langle x^m (t) \rangle.
\end{equation}
Thus, in the limit $t\to\infty$, we obtain an exact representation of the Fourier-transformed propagator:
\begin{equation}
\widehat{W}(k,t \to \infty) =\sum_{m=0}^{\infty} \frac{(-1)^m k^{2m}}{[(2m)!!]^{\alpha}} \mathfrak{D}_{\alpha}^m H^{- m \alpha} \equiv \sum_{m=0}^{\infty} c_m Z^{2m},
\end{equation}
where $c_m = {(-1)^m }/{[(2m)!!]^{\alpha}}$ and $Z = k \mathfrak{D}_{\alpha}^{{1}/{2}} H^{ {- \alpha}/{2}}$.

Unfortunately, the sum $\sum_{m=0}^{\infty} c_m Z^{2m}$ is only known for the case $\alpha=1$ (normal diffusion), which yields $\widehat{W}(k,t \to \infty) =\exp(-Z^2/2)$. In this case, the inverse Fourier transform is known, and one obtains
\begin{equation}
W(x,t \to \infty) = \frac{1}{\sqrt{2 \pi {\mathfrak{D}}/{H}}} \exp \left(- \frac{x^2}{{2 \mathfrak{D}}/{H}} \right),
\end{equation}
with $\mathfrak{D} \equiv \mathfrak{D}_1$. As already mentioned, for $\alpha \neq 1$ there is no known analytic expression for $\sum_{m=0}^{\infty} c_m Z^{2m}$, and one must resort to the numerical inversion of this quantity to obtain a semianalytical result for the comoving propagator.
The reconstruction of $W(x,\infty)$ from its moments given by Eq.~\eqref{xinftMoments} could be an alternative way \cite{John07}, but it is beyond the scope of this paper.

\subsubsection{Exponential expansion: moments in physical space}

\begin{figure}[t]
 {\includegraphics[width=0.5\textwidth]{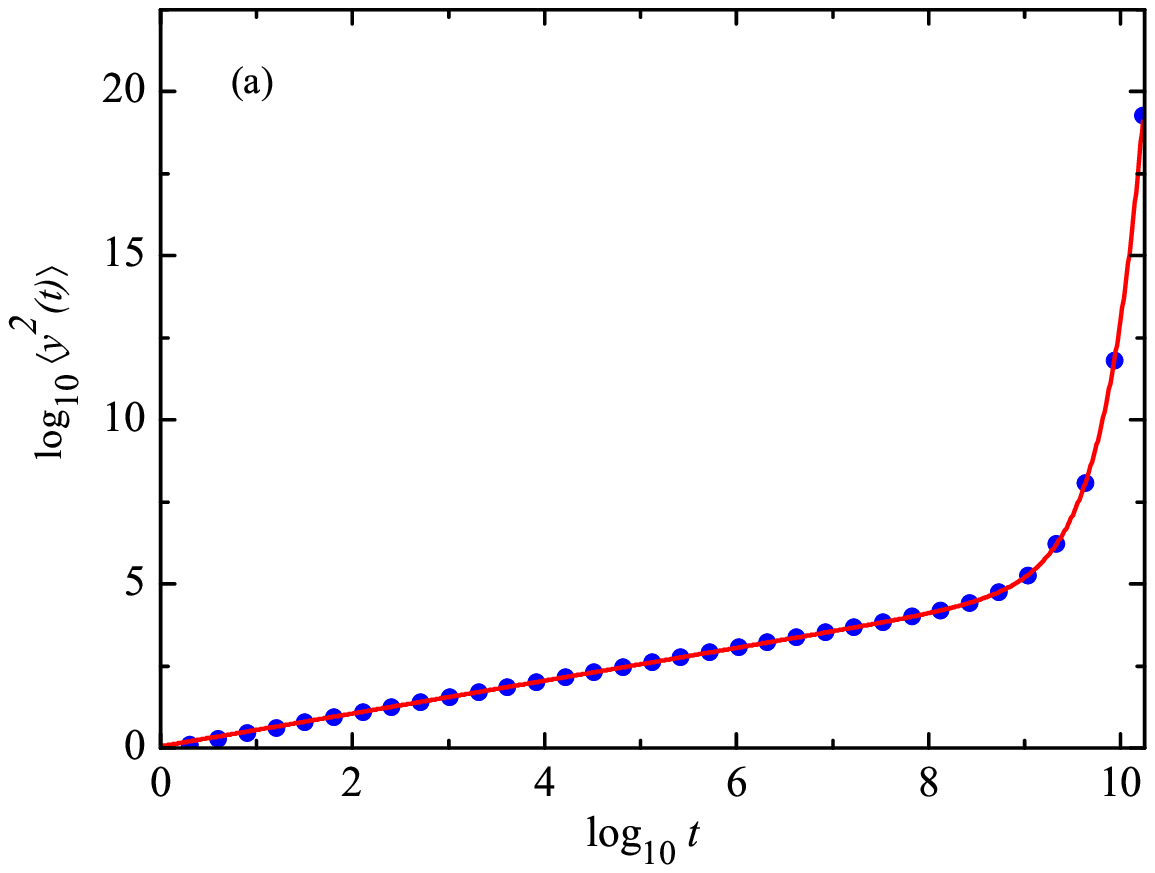}}
 {\includegraphics[width=0.5\textwidth]{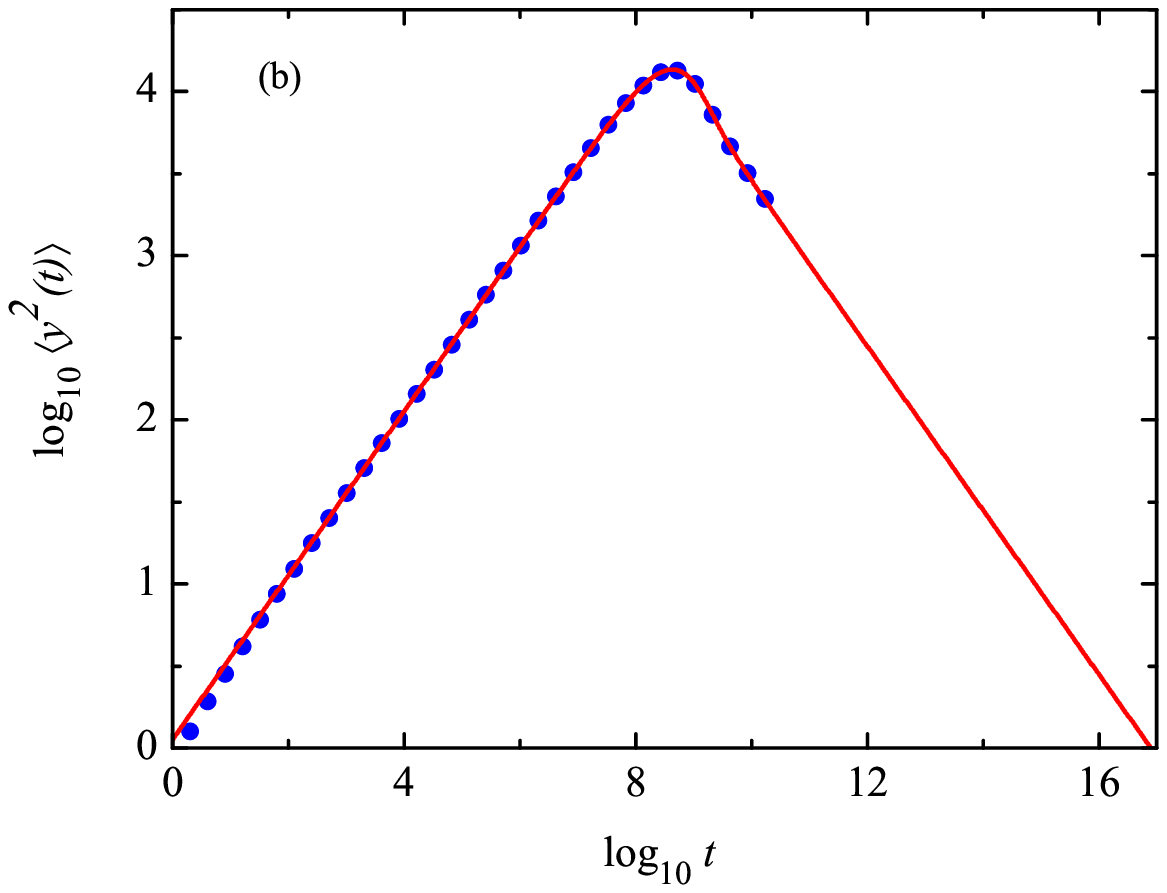}}
\caption{Double-logarithmic representation of the physical second order moment corresponding to a subdiffusive random walk in an exponentially expanding medium with (a) $H=10^{-9}$ and (b) $H=-10^{-9}$ (contracting medium). In all cases $\alpha=1/2$ and $\mathfrak{D}_{\alpha}=1/2$. Solid lines represent analytical results obtained from Eq.~(\ref{SecondOrderComovingMomentumExpExp}). The dots correspond to numerical simulations performed up to $t=2^{34}$ with $10^6$ realizations  where $\lambda(y)$ is given by  Eq.~\eqref{gaussianLambda1} with $\sigma^2=1/2$ and $\varphi(t)$ is given by Eq.~\eqref{paretoSimu}  with $\xi=1/\pi$. For the contracting medium, panel (b), one sees that $\langle y^2 \rangle$ goes to zero at sufficiently long times. }
\label{Fig:ProperSecondMomentSubdiffusionExpExp}
\end{figure}

From Eq.~\eqref{SecondOrderComovingMomentumExpExp} one has
\begin{equation}
\langle y^2 (t) \rangle =a^2(t) \langle x^2 (t) \rangle =  {2 \mathfrak{D}_{\alpha}} (2H)^{- \alpha} \exp(2Ht) \frac{\gamma(\alpha,2Ht)}{ \Gamma(\alpha)} .
\label{y2expo}
\end{equation}
The lower incomplete gamma function becomes the gamma function for $Ht\to\infty$ so that, for long times,
\begin{equation}
\langle y^2 (t) \rangle \sim  2 \mathfrak{D}_{\alpha} (2H)^{-\alpha} \exp(2Ht).
\end{equation}
This long-time exponential growth of the second-order moment is depicted in Fig.~\ref{Fig:ProperSecondMomentSubdiffusionExpExp}a. This is a signature of the fact that the physical propagator widens exponentially, both in the normal diffusion case ($\alpha=1$) and in the anomalous diffusion case ($\alpha <1$), that is, the spreading due to diffusion is subdominant with respect to the displacement induced by the exponential expansion of the medium.

\subsubsection{Exponential contraction. The Big Crunch}
\label{secBigCrunch}

In the case of an exponentially contracting medium ($H<0$), the behavior of the physical moments displays interesting features with a markedly different behavior for normal diffusion $\alpha=1$ and for anomalous diffusion $0<\alpha<1$. As we shall see, a minute amount of anomalous diffusion introduces a qualitative change in the behavior of the propagator.

In the normal diffusion case $\alpha=1$ the propagator in physical space can be shown to converge to a ``frozen'' profile for $H<0$.
The stationarity of the long-time propagator implies that the time growth of each of the corresponding even moments is slowed down at long times and eventually a plateau is reached. Since in the present case of normal diffusion the stationary propagator is Gaussian, the integrals corresponding to the associated moments are well-known and an explicit evaluation of the latter becomes possible. It is nevertheless instructive to follow an alternative derivation based on the hierarchy of differential equations connecting time derivatives of higher order moments with lower order moments. More specifically, let us consider the differential equation for the second moment in physical space (see \cite{Yuste2016}), i.e.,

\begin{equation}
\frac{d \langle y^2 (t) \rangle}{dt} - 2 H  \langle y^2 (t) \rangle = 2 \mathfrak{D}.
\end{equation}
This is a linear first-order differential equation whose solution reads
\begin{equation}
 \langle y^2 (t) \rangle = \frac{ \mathfrak{D}}{H} \left[ \exp(2Ht) - 1 \right].
\end{equation}
As expected, in the long-time regime this function converges to the constant $\mathfrak{D}/|H|$ when $H<0$.

However, the behavior changes dramatically as soon as $\alpha<1$.  As we have seen, in the $H>0$ case there is no qualitative difference with respect to the normal diffusion case. In contrast, in the present case of a contracting medium ($H<0$) all the moments in physical space will eventually collapse to zero. In Cosmology, this eventual singularity is called ``The Big Crunch".

 Despite the eventual collapse of all the moments, the behavior is non-monotonic in time, as exemplified by the time dependence of the second-moment depicted in Fig.~\ref{Fig:ProperSecondMomentSubdiffusionExpExp}b. In this case, the second moment $\langle y^2 (t) \rangle $ is also given by \eqref{y2expo}.  For $H<0$, this moment  increases in time until a maximum or ``turning point", and from then on its value decreases to zero.  The long time behavior can   be obtained from the asymptotic properties of gamma functions:
\begin{equation}
\langle y^2 (t) \rangle   \sim  - \frac{\mathfrak{D}_{\alpha} }{H \Gamma(\alpha)} \, t^{\alpha-1}
\end{equation}
As it turns out, the $t^{\alpha-1}$-behavior displayed by the second-order moment in the long time limit for $H<0$ extends to the entire set of even moments (odd moments vanish because of the absence of drift). As a result of this, the propagator in physical space tends to $\delta(y)$. In order to prove this result, we shall first show that  the Laplace transformed ($2m^{\text{th}}$-order moment) in physical space is given by the following expression:
\begin{equation}
\langle y^{2m} (s) \rangle = (2m)! \mathfrak{D}_{\alpha}^{m} (s-2mH)^{-1} \prod_{n=0}^{m-1} (s-2nH)^{-\alpha}, \quad m=1,2...
\label{genexpym}
\end{equation}
The proof is elementary and based on mathematical induction with respect to the order $m$. The case $m=2$ is immediately given by Eq.~(\ref{mthPhysicalMomentLaplace}b). Then one assumes that Eq.~\eqref{genexpym} holds for $m-2$ and insertion into Eq.~(\ref{mthPhysicalMomentLaplace}a) concludes the proof.

Next, we make use of  Eq.~\eqref{genexpym} to obtain the behavior in the $s \to 0$ limit. We obtain
\begin{equation}
\langle y^{2m} (s) \rangle \sim (2m-1)! \mathfrak{D}_{\alpha}^m \left[ (2m-2)!! \right]^{-\alpha} |H|^{\alpha \left( 1 - m \right) -1} s^{-\alpha}, \quad m=1,2,\ldots
\end{equation}
which, as already anticipated, corresponds to the long time behavior
\begin{equation}
\langle y^{2m} (t) \rangle \sim \frac{1}{\Gamma (\alpha)} (2m-1)!\, \mathfrak{D}_{\alpha}^m \,\left[ (2m-2)!! \right]^{-\alpha} |H|^{\alpha \left(1-m \right) -1}\, t^{\alpha -1}, \quad m=1,2,\ldots
\end{equation}
We thus see that, as soon as $\alpha < 1$, all the physical moments tend to zero  and a Big Crunch takes place. In contrast, for $\alpha = 1$ all the moments tend to a constant value, which leads to a stationary propagator \cite{Yuste2016}.

How can we intuitively understand the above behavior? In contrast to the normal diffusion case, when $\alpha<1$ aging effects come into play, and the average jump rate decreases monotonically in time \cite{Klafter2011, Soko2012, Metzler2014b}. Thus, as time goes by, particle jumps become less frequent. This results in less efficient particle spreading and, ultimately, in the inability to compensate the drift towards the origin induced by the contraction.

So far, we have focused on the decay of the positional moments in the long-time regime. It is, however, remarkable that in the opposite limit of sufficiently short times the effect of the medium contraction is hardly noticeable, and diffusive spreading dominates. Thus, there exists a crossover time between a short-time regime where a moment of a certain order grows, and a long-time regime where it decreases and eventually tends to zero. This crossover time, which we term``turning time'' in what follows, corresponds to the instant where a given moment reaches its maximum value in physical space (turning point). The turning time turns out to depend weakly on the order of the moment. For our purpose here, which is a rough characterisation of the typical time scale associated with the onset of the Big Crunch effect, it will suffice to compute the turning time $t_T$  corresponding to the second-order moment $\langle y^2(t) \rangle$. The location of $t_T$ on the time axis can be computed from the condition that the time derivative $\left.{d \langle y^2 (t) \rangle}/{dt}\right|_{t=t_T}$ must vanish. This yields
 \begin{equation}
   z^{1-\alpha} \exp(z) \gamma (\alpha,z) +1 =0,
 \label{zEq}
\end{equation}
with $z=2t_TH$. For each value of $\alpha$, this is a transcendental equation whose unique solution $z^*(\alpha)$ can be found numerically. The corresponding turning time  is $t_T=z^*(\alpha)/(2H)$.  For example, for $\alpha=1/2$ one finds $z^*=-0.854$. For
$H=-10^{-9}$ and $\mathfrak{D}_{\alpha}=1/2$ (values chosen in Fig.~\ref{Fig:ProperSecondMomentSubdiffusionExpExp}b) one has $t_T =4.27 \times 10^8$ and $\langle y^2 (t_T) \rangle = 1.37\times 10^4$. This result, along with the analytical prediction for the full time evolution, is confirmed by numerical simulations (see Fig.~\ref{Fig:ProperSecondMomentSubdiffusionExpExp}b). Equation \eqref{zEq} lacks a solution for $\alpha\ge 1$ or for $z \geq 0$, in agreement with the fact that there is no finite turning time for normal diffusion as well as for a non-contracting medium.

\section{Conclusions}
\label{Conclusions}

In this work we derived a fractional diffusion equation for a broad class of diffusion processes described by a CTRW model. The dynamics of this model takes place in a uniformly expanding (contracting) medium. The derivation of the anomalous diffusion equation draws on the key observation that, upon switching to comoving coordinates, the relevant integral equations adopt the same form as the ones for the standard case of a static medium, except for the fact that the jump length pdf becomes time dependent. In spite of this time dependence introduced by the medium expansion, standard tools for the derivation of fractional differential equations still apply and can be used to derive a very general bifractional diffusion equation.

Upon deriving the diffusion equation, we went on to study some special, yet representative cases such as L\'evy flights and subdiffusive CTRWs.
For L\'evy flights with an exponential waiting time pdf, the propagator was straightforwardly obtained from the standard solution in a static medium by replacing the time variable with a so-called diffusion conformal time $T(t)$ describing the effect of the medium expansion (contraction). As it turns out, for $t\to\infty$ the diffusion conformal time $T$ may either tend to infinity or saturate to a finite value. In the latter case, a strong modification of the behavior of the moments as well as of the propagator in physical space is observed. For instance, in the case of a L\'evy flight in an exponentially contracting medium, the long-time pdf of the particle's position approaches a stationary profile in physical space. Another interesting consequence is that, for a sufficiently fast medium expansion, the mixing of diffusive particles initially separated by a comoving distance significantly larger than the ``L\'evy event horizon'' will remain negligible at all times. Likewise,  after a finite time $t$ there will be no significant mixing provided that the initial separation is significantly larger than the ``L\'evy horizon'', that is, the typical comoving distance traveled by the diffusing particles up to time $t$.

As regards subdiffusive CTRWs in an expanding (contracting) medium,
by switching to comoving coordinates, the relevant diffusion equation could be cast into a fractional diffusion equation with a time-dependent diffusion coefficient.
In the case of an exponential expansion we were able to compute the long-time behavior of all the moments. For an exponentially contracting medium, we observed the onset of a remarkable ``Big Crunch'' effect induced by the long tail of the waiting time pdf. This effect originates from inefficient particle spreading due to ``aging'' phenomena characteristic of the subdiffusive CTRW model, and it is totally absent in the normal diffusion case.

The above results may be useful in different fields, e.g., in Cosmology. Fractional diffusion equations have indeed already been put forward to address some anomalous diffusion problems related to photon scattering in inhomogeneous magnetic fields, an scenario somewhat reminiscent of the diffusion of cosmic rays \cite{Lagutin2003, Uchaikin2013}. In this context, Berezinsky and Gazizov \cite{Berezinsky2006} have pointed out that the characteristics of cosmic ray propagation in extragalactic space depend on the specific region they traverse (voids, filaments, and clusters of galaxies). According to these authors, diffusive transport occurs only at energies for which the diffusion length is smaller than the typical linear size of the region, whereas the propagation is quasi-ballistic in the opposite limit. This transition between a normal diffusive regime and a ballistic regime (presumably via an intermediate superdiffusive regime) highlights the potential interest of considering anomalous diffusion problems in the context of Cosmology, where medium expansion effects appear in a very natural way.

Finally, we note that our results may also be relevant in a biological context. More specifically, tissue growth has been suggested to occur on time scales commensurate with the formation of the Dpp morphogen gradient \cite{Fried2014,
Averbukh2014}. On the other hand, the dissemination of this morphogen occurs via different pathways \cite{Ibanes2008}, and it is therefore unlikely that a simple picture based on Fickian diffusion provides an accurate description of the gradient formation. Therefore, anomalous diffusion models are expected to play an important role here \cite{Hornung2005, Kruse2008, YusteAbadLindenberg2010, Boon2012, Yin2013, FedotovFalconer2014}. In this context, one should not forget that morphogens are subject to degradation processes \cite{Ibanes2008}. Fortunately, CTRW models lend themselves particularly well to the inclusion of degradation and, more generally, of chemical reactions \cite{VladRoss02, Mendez2004, HenryLanglandsWearnePRE06, SokolovSchmidtSaguesPRE06,
YadavHorsthemke06, SekiJPCM07, FroembergSokolovPRL08, HenryLanglandsWearnePRE08, Campos2008, AYL2010, MFHbook,
Fedotov2010, YALBookChapter2011, AYL2012, AYL2013, AYL2013b, Angstmann2013, AYLBookChapter2014, Campos2015, Angstmann2016}.
Given the structure of the CTRW equations in the comoving space, we expect that the inclusion of simple chemical reactions can be carried out in a similar way as it is done in the case of a static medium. Research concerning both CTRW models with chemical reactions and the specific case of morphogen gradient formation is currently underway.

\section{Acknowledgments}

This work was partially funded by MINECO (Spain) through Grants No. FIS2016-76359-P (partially financed by FEDER funds) (S.~B.~Y. and E.~A.) and by the Junta de Extremadura through Grant No. GR15104 (S.~B.~Y. and E.~A.). F. ~L.~V. acknowledges generous financial support from the Fundaci\'on Tatiana P\'erez de Guzm\'an El Bueno and from the Junta de Extremadura through Grant. No. PD16010 (FSE funds).

\appendix*
\section{Simulation procedure}

 The CTRW model has been used over several decades to mimic anomalous diffusion processes in static media, e.g. L\'evy flights \cite{Abdel-Rehim2008}. The standard computational algorithm used to implement the dynamics defined by CTRW model is relatively straightforward. In what follows, we show how to extend this algorithm to the case of an expanding (or contracting) medium.

 As explained in the main text, a walker chosen arbitrarily from an ensemble is assumed to be at a certain physical point $y_0$ at some initial time $t_0$. At this initial time, the comoving distance and physical distance are identical. As the medium grows, the walker's physical position $y$ changes in time due to two contributions, namely, a continuous drift arising from the Hubble flux, and the intrinsic displacement due to occasional instantaneous discrete jumps. The complication introduced by the drift due to the medium expansion (contraction) can be circumvented by exploiting the fact that the walker's comoving position  $x$ does not change as long as the walker does not jump.

 Denoting by $t_n$ the time at which the walker performs the $n$th jump, one has
 \begin{equation}
 t_{n+1}=t_n + \Delta t
\end{equation}
 where the waiting time $\Delta t$ is drawn from the waiting time pdf $\varphi(\Delta t)$. In this article,  $\varphi(\Delta t)$ has been taken to be exponential [Eq.\eqref{expphi}]  for L\'evy flights (Figs.~\ref{FigPHm001} and \ref{Fig:LevyPulsesComovingPropagators})   and a Pareto distribution [Eq.\eqref{paretoSimu}] for subdiffusive CTRWs (Figs.~\ref{Fig:ComovingSecondMomentSubdiffusionPotentialExp} and \ref{Fig:ProperSecondMomentSubdiffusionExpExp}).

 Denoting by $t$ an arbitrary intermediate time between two consecutive jumps ($t_n < t \le t_{n+1}$), the corresponding physical position will be \begin{equation}
 \label{ynxn}
 y(t) = a(t) x_n
 \end{equation}
 where $x_n=y(t_n)/ a(t_n)$ and $y(t_n)$ stand for the comoving and physical coordinates, respectively, immediately after the $n$th jump.
Because the walker's comoving position  $x$ does not change as long as the walker does not jump, one has $x(t)=x_n$ for $t_n<t\le t_{n+1}$.
Therefore, if the walker takes a jump of length $\Delta y$ at time $t_{n+1}$, then the  position of the walker immediately after  this  $n+1$th step is
  \begin{equation}
\label{yjump}
 y(t_{n+1}) = \frac{a(t_{n+1})}{a(t_n)} y(t_n)  + \Delta y.
 \end{equation}
The jump displacement $\Delta y $ is drawn from the jump length pdf $\lambda(\Delta y)$.   Equations  \eqref{ynxn} and \eqref{yjump} provide the physical position of the walker at any time.  In the case of symmetric L\'evy flights,   $\Delta y$ has been generated by means of the Chambers-Mallows-Stuck method \cite{Klafter2011,Abdel-Rehim2008,Chambers1976}. On the other hand, the simple inversion method of the cumulative distribution function \cite{Klafter2011} is used for the particular case of the Cauchy distribution.

\end{document}